\documentclass[preprint]{aastex6}
\newcommand{\nraoblurb}{The National Radio Astronomy Observatory is
a facility of the National Science Foundation operated under cooperative
agreement by Associated Universities, Inc.}
\newcommand{\degree}{\ensuremath{\,^\circ}}

\newcommand{\arcsper}{\rlap.{^{\prime\prime}}}

\newcommand{\mhz}{\ensuremath{\,{\rm MHz}}}
\newcommand{\ghz}{\ensuremath{\,{\rm GHz}}}

\newcommand{\K}{\ensuremath{\,{\rm K}}}

\newcommand{\mm}{\ensuremath{\,{\rm mm}}}

\newcommand{\percc}{\ensuremath{\,{\rm cm^{-3}}}}
\newcommand{\Mpc}{\ensuremath{\,{\rm Mpc}}}

\newcommand{\pc}{\ensuremath{\,{\rm pc}}}
\newcommand{\kms}{\ensuremath{\,{\rm km\, s^{-1}}}}
\newcommand{\msun}{\ensuremath{\,M_\odot}}
\newcommand{\sfr}{\ensuremath{\,M_\odot\,{\rm yr}^{-1}}}
\newcommand{\s}{\,s}
\newcommand{\hr}{\,hr}
\newcommand{\microns}{\ensuremath{\, \mu {\rm m}}}

\newcommand{\mjy}{\,mJy}
\newcommand{\mjyb}{\ensuremath{\rm \,mJy\,beam^{-1}}}

\newcommand{\microjyb}{\ensuremath{\rm \,\mu Jy\,beam^{-1}}}

\newcommand{\te}{\ensuremath{T_{\rm e}}}

\newcommand{\Ne}{\ensuremath{n_{\rm e}}}

\newcommand{\hrrl}[1]{H{#1}$\alpha$}

\newcommand{\nexpo}[2]{\ensuremath{#1 \times 10^{#2}}}

\newcommand{\hii}{H\,{\sc ii}}
\newcommand{\cii}{C\,{\sc ii}}
\newcommand{\nii}{N\,{\sc ii}}

\newcommand{\ic}[1]{IC\thinspace #1}

\newcommand{\gsim}{\ensuremath{\gtrsim}}
\newcommand{\lsim}{\ensuremath{\lesssim}}
\newcommand\urltilda{\kern -.15em\lower .7ex\hbox{\~{}}\kern .04em}

\begin{document}

\title{JVLA Observations of \ic342: Probing Star Formation in the Nucleus}


\author{Dana S. Balser\altaffilmark{1}, Trey V. Wenger\altaffilmark{1,2}, W. M. Goss\altaffilmark{3}, 
K. E. Johnson\altaffilmark{2}, \& Amanda  A. Kepley\altaffilmark{1}}

\altaffiltext{1}{National Radio Astronomy Observatory, 520 Edgemont Rd., Charlottesville, VA 22903, USA.}
\altaffiltext{2}{Department of Astronomy, University of Virginia, 530 McCormick Road, Charlottesville, VA 22903, USA.}
\altaffiltext{3}{National Radio Astronomy Observatory, P.O. Box 0, Socorro, NM 87801, USA.}


\begin{abstract}

  \ic{342} is a nearby, late-type spiral galaxy with a young nuclear
  star cluster surrounded by several giant molecular clouds.  The
  \ic{342} nuclear region is similar to the Milky Way and therefore
  provides an interesting comparison.  We explore star formation in
  the nucleus using radio recombination line (RRL) and continuum
  emission at 5, 6.7, 33, and 35\ghz\ with the JVLA.  These radio
  tracers are largely unaffected by dust and therefore sensitive to
  all of the thermal emission from the ionized gas produced by
  early-type stars.  We resolve two components in the RRL and
  continuum emission within the nuclear region that lie east and west
  of the central star cluster.  These components are associated both
  spatially and kinematically with two giant molecular clouds.  We
  model these regions in two ways: a simple model consisting of
  uniform gas radiating in spontaneous emission, or as a collection of
  many compact \hii\ regions in non-LTE.  The multiple \hii\ region
  model provides a better fit to the data and predicts many dense
  ($n_{\rm e} \sim 10^{4}-10^{5}$\percc), compact ($\lsim 0.1$\pc)
  \hii\ regions.  For the whole nuclear region as defined by RRL
  emission, we estimate a hydrogen ionizing rate of $N_{\rm L} \sim$
  \nexpo{2}{52}$\,{\rm s}^{-1}$, corresponding to equivalent $\sim
  2000$ O6 stars and a star formation rate of $\sim 0.15$\sfr.  We
  detect radio continuum emission west of the southern molecular mini
  spiral arm, consistent with trailing spiral arms.

\end{abstract}

\keywords{galaxies: individual (\ic{342}) --- galaxies: ISM ---
  galaxies: star formation --- radio continuum: galaxies --- radio
  lines: galaxies}

\section{Introduction}

Young massive star clusters are found in starburst galaxies, normal
galaxies, galactic nuclei, the central regions of mergers, and in
tidal tails \citep{boker04,degrijs04,knierman04,larsen04}.  These star
clusters have many names such as super star clusters,
super-associations, starbursts, etc.  The most massive star clusters
in the Galaxy reside in the optically obscured Galactic Center region
and are small compared to those found in the local Universe.  The
nearest prototype for these larger star clusters is 30 Doradus in the
Large Magellanic Cloud \citep{ambart54}.  The {\it Hubble Space
  Telescope (HST)} has been pivotal in discovering more distant star
clusters
\citep{holtzman92,johnson00,whitmore03,leitherer03,oconnell04}.  The
most dense and massive young star clusters appear to be young versions
of classical globular clusters and are an important mode of star
formation and evolution.

Very young star clusters are difficult to observe in the UV, optical,
or near IR because of the high extinction from their surrounding natal
clouds of gas and dust.  Radio continuum measurements, however, are
excellent probes of these regions where observations at several
frequencies can separate the thermal emission, associated with HII
regions, and the non-thermal emission, related to supernova remnants
\citep{turner83,kobulnicky99,turner00,johnson03,johnson04}.  The
measured Lyman continuum fluxes provide an estimate of the OB stellar
luminosities that are typically consistent with far-infrared
luminosities \citep{turner94}.  Radio recombination lines (RRLs) are
more difficult to detect but yield important information about the
dynamics of the brightest, nearby star clusters
\citep{anantharamaiah93,anantharamaiah96,anantharamaiah00,mohan02}.
Young massive star clusters have now been detected in about a dozen
nearby galaxies in RRL emission, primarily with interferometers where
the high spatial resolution can resolve individual clusters for nearby
galaxies \citep[e.g.,][]{rodriguez05}; otherwise we are probing the
bulk emission over a larger region within the galaxy.  Both models and
observations show that RRL emission is more intense at higher
frequencies primarily due to free-free opacity.  For example, the 43
GHz RRL emission towards Arp220 is 50 times more intense than the 8
GHz RRL emission \citep{rodriguez05}.

Because RRLs are weaker than optical tracers their use has been
limited to only the brightest star forming regions.  Improvements in
frequency coverage and sensitivity of the Very Large Array, however,
have increased the sensitivity to RRL work by almost an order of
magnitude \citep{kepley11}.  At lower frequencies many RRLs can be
observed {\it in parallel} and averaged to improve the sensitivity for
these weaker RRL transitions, and the wider bandwidths provides
sufficient coverage to better detect higher frequency RRLs.
Therefore, less massive star clusters can now be detected in RRL
emission.

Here we focus on the nearby, face-on, late-type spiral galaxy \ic{342}
which contains a nuclear star cluster.  Observations reveal that many
of the properties of the nucleus of \ic{342} are similar to the
Galactic Center \citep{meier14}.  For example, the size and mass of
the central molecular zones and ionizing photon rates.  Therefore,
understanding the nuclear star cluster in \ic{342} can shed light on
star formation in the Galactic Center, where our position in the disk
together with dust obscuration limits our view.

\begin{deluxetable}{lll}
\tablecaption{JVLA Observational Setup \label{tab:obs}}
\tablehead{
\colhead{Parameters} & \colhead{Ka-band} & \colhead{C-band}
}
\startdata
Proposal Code           & 11B-078                  & 12A-186 \\
Observing Dates         & 2011 Nov 27, Dec 8, 31   & 2012 Mar 17; 2012 Jan 1,2  \\
Configuration           & D                        & C \\
Phase Calibrators       & J0228+6721               & J0228+6721 \\    
Flux Density Calibrators & 0542+498 (3C147)         & 0137+331 (3C48) \\
Bandpass Calibrators    & J0319+4130               & J0319+4130 \\
Time on Source (hr)     & 5.0                      & 3.5 \\
Sky Frequency (GHz)\tablenotemark{a}     & 34.65681694, 32.91281657 & 4.99214252, 6.72684455 \\
Continuum BW (MHz)      & 768.0                    & 1024.0 \\
Line BW (MHz)           & 128.0                    & 16.0 \\ 
Line BW (\kms)\tablenotemark{a}          & 1160.0, 1160.0           & 960.0, 711.0 \\
Channel Width (MHz)     & 1.21                      & 0.0756 \\
Channel Width (\kms)\tablenotemark{a}    & 10.5, 11.0                 & 4.5, 3.4 \\
Number of Channels      & 128                      & 256 \\
\enddata
\tablecomments{\ic{342} position: RA(J2000) = 03:46:48.514; Dec(J2000) = 68:05:45.98} 
\tablenotetext{a}{The two numbers correspond to the two basebands.}
\end{deluxetable}

\section{JVLA Observations and Data Reduction}\label{sec:obs}

Here we use the National Radio Astronomy Observatory
(NRAO)\footnote{\nraoblurb} Jansky Very Large Array (JVLA) to observe
the RRL and continuum emission at C-band (5 and 6.75\ghz) and Ka-band
(32 and 34\ghz) toward the nuclear region of \ic{342}.  RRLs provide
information about the electron density and temperature along with the
number of hydrogen ionizing photons emitted nearby early-type stars
\citep[e.g.,][]{shaver75, balser99} These diagnostics therefore are an
unobscured tracer of high mass star formation.  Typically, to
constrain \hii\ region models at least two RRLs separated in frequency
are required \citep[e.g.,][]{anantharamaiah93, balser16}.
Table~\ref{tab:obs} summarizes the JVLA observations.  The JVLA
configurations were chosen to provide similar spatial sampling between
the two frequency bands.  Table~\ref{tab:transitions} lists the
observed RRL transitions.  There are only two RRLs in the 2\ghz\
sampled at Ka-band separated by about 1.7\ghz.  At C-band we observed
one set of 5 RRLs near 6.75\ghz\ and another set of 8 RRLs near
5.0\ghz.

\begin{deluxetable}{lcc}
\tablecaption{RRL Transitions \label{tab:transitions}}
\tablehead{
\colhead{} & \colhead{Rest} & \colhead{Baseband} \\
\colhead{} & \colhead{Frequency} & \colhead{Frequency} \\
\colhead{Name} & \colhead{(GHz)} & \colhead{(GHz)} 
}
\startdata
\multicolumn{3}{c}{\underline{Ka-band}} \hfill \\
\hrrl{57}   & 34.5964 &  34.6568 \\ 
\hrrl{58}   & 32.8522 &  32.9128 \\
\multicolumn{3}{c}{\underline{C-band}} \hfill \\
\hrrl{97}   &  7.0954 &   6.7268  \\
\hrrl{98}   &  6.8815 &   6.7268  \\
\hrrl{99}   &  6.6761 &   6.7268  \\
\hrrl{100}  &  6.4788 &   6.7268  \\
\hrrl{101}  &  6.2891 &   6.7268  \\
\hrrl{106}  &  5.4443 &   4.9921  \\
\hrrl{107}  &  5.2937 &   4.9921  \\
\hrrl{108}  &  5.1487 &   4.9921  \\
\hrrl{109}  &  5.0089 &   4.9921  \\
\hrrl{110}  &  4.8742 &   4.9921  \\
\hrrl{111}  &  4.7442 &   4.9921  \\
\hrrl{112}  &  4.6188 &   4.9921  \\
\hrrl{113}  &  4.4978 &   4.9921  \\
\enddata
\end{deluxetable}

\subsection{Ka-band (33 and 35\ghz)}\label{sec:obs-kaband}

The correlator was configured in Open Shared Risk Observing mode to
provide two 1\ghz\ basebands. Each baseband had eight, 128\mhz-wide
sub-bands. The \hrrl{57} and \hrrl{58} lines were placed in the fourth
sub-band of the first and second basebands, respectively. The other 7
sub-bands in each baseband measured the continuum emission.  We used
the Common Astronomy Software Applications (CASA)\footnote{See
  http://casa.nrao.edu} package to reduce the JVLA data.  Observations
at these frequencies requires that the observed amplitudes be
corrected for atmospheric opacity and the telescope gain as a function
of elevation. We used the seasonal average to determine the opacity
for observations between 23 December 2011, 19:12 UT to 12 January
2012, 23:37 UT because the weather station was not working between
those dates (G. van Moorsel, private communication, 2012). We took the
average of the seasonal and measured opacity as the true opacity for
the rest of the observations. The gain response as a function of
frequency was corrected with the standard telescope gain curves in
CASA 4.0.1.

The data were calibrated using the following procedure.  First we
created a model image of the bandpass calibrator.  The flux density
and bandpass calibrator phases and amplitudes were determined per
integration time (3\s) and then corrections were made.  Estimates of
the bandpass were derived using the flux density calibrator and a
polynominal bandpass solution. The estimated bandpasses were applied
to the data and the flux density and bandpass calibrator phases and
amplitudes determined per integration time. The flux density scale for
the bandpass calibrator was determined using the flux density
calibrator and the Perley-Butler 2010 flux density models
\citep{perley13}.  Next, an image of the bandpass calibrator was
created and used to derive the final bandpasses.  As before, the
bandpass and complex gain calibrators have their phases and amplitudes
corrected per integration (3\s). The bandpass was then determined on a
per-channel basis using the bandpass calibrator. The final bandpasses
were applied and the bandpass and complex gain calibrator phases and
amplitudes determined on a per-integration time and a per-scan
basis. The per-scan phases and amplitudes from the complex gain
calibrator were applied to the source data. The flux density of the
source data was derived using the model of the bandpass calibrator
determined in the first part of the calibration.

After the calibration step, the continuum data for the target were
combined to form two final continuum data sets at 33\ghz\ and
35\ghz. These data were phase and amplitude self-calibrated.  These
solutions were applied to the sub-bands containing the line emission.

\subsection{C-band (5 and 6.7\ghz)}\label{sec:obs-cband}

The correlator was configured in Resident Shared Risk
Observing mode. Two 1\ghz\ wide basebands (each with eight 128\mhz\
wide sub-bands) were used to measure the continuum at 5.0 and
6.7\ghz. Thirteen 16\mhz\ wide sub-bands were tuned to RRLs in the
1\ghz\ available for each baseband. The lines observed were
\hrrl{97}-\hrrl{101} at 6.7\ghz\ and \hrrl{106}-\hrrl{113} at
5.0\ghz.  Interference is significant at these frequencies, in
particular at 6.75\ghz. The data were flagged automatically using the
RFLAG algorithm (Appendix E in \citet{greisen12}). Channels containing
particularly egregious interference were flagged outright for the
entire data set.

The calibration for the continuum data followed a similar procedure to
that for the Ka-band data (\S~\ref{sec:obs-kaband}). The phases and
amplitudes for the bandpass and flux density calibrator were
determined per integration as before. The phases and amplitudes for
the complex gain calibrator, however, were determined on a per-scan
basis. The line data were calibrated similarly to the continuum data
except for one significant difference.  We used the continuum data to
generate a model for the bandpass calibrator instead of the line data
since the wider bandwidths of the continuum data produced a better
model of the bandpass.  This also reduced the time to generate the
calibration since only one iteration over the line data was necessary.
The bandpass stability for both the line and continuum data was
checked using observations of the bandpass calibrator taken every
1.5\hr\ during the observing run. A few antennas with time variable
bandpasses were flagged.  To improve the complex gain solutions, the
continuum data sets were phase and amplitude self-calibrated. The
solutions from the continuum self-calibration were applied to the line
data.

\subsection{Imaging}\label{obs:imaging}

Continuum images were created using multi-frequency synthesis and
cleaned using multi-scale clean.  We adopted a minimum {\it u-v}
distance to ensure that all images have the same largest angular
scale; this distance corresponds to the largest angular scale sampled
by the highest frequency continuum data.  We also smoothed the images
to a common spatial resolution of $4\arcsper5$, which corresponds to
the angular resolution of the lowest frequency line data.  

Spectral line cubes were continuum subtracted in the {\it u-v} space
and then imaged using the same spatial scale as the continuum images.
We used a uniform velocity resolution of 20\kms\ to maximize the
signal-to-noise ratio while still resolving the velocity structure of
the lines.  For the C-band data, the individual spectral line cubes
were averaged together to create two ``stacked'' spectral line cubes
near 5 and 6.7\ghz.  RRL transitions occur at large principal quantum
numbers (n $ > 50$), and therefore only small variations exist in the
peak intensity between adjacent Hn$\alpha$ transitions.  Here we do
not correct for these differences when stacking RRLs.

\section{Results}

Our goal was to measure the ionizing flux and star formation rate at
the center of \ic{342} using RRL and continuum emission at four
frequencies: 5\ghz\ (\hrrl{106}-\hrrl{113}), 6.7\ghz\
(\hrrl{97}-\hrrl{101}), 34\ghz\ (\hrrl{58}), and 35\ghz\ (\hrrl{57}).
This provides four independent data points to constrain a variety of
models.  We chose these frequencies to provide maximum leverage for
our models based on the results of previous studies
\citep[e.g.,][]{anantharamaiah00}.  We need to average (stack) the
RRLs at C-band since these lower frequency RRLs are weaker.

\begin{figure}
\includegraphics[angle=0,scale=0.8]{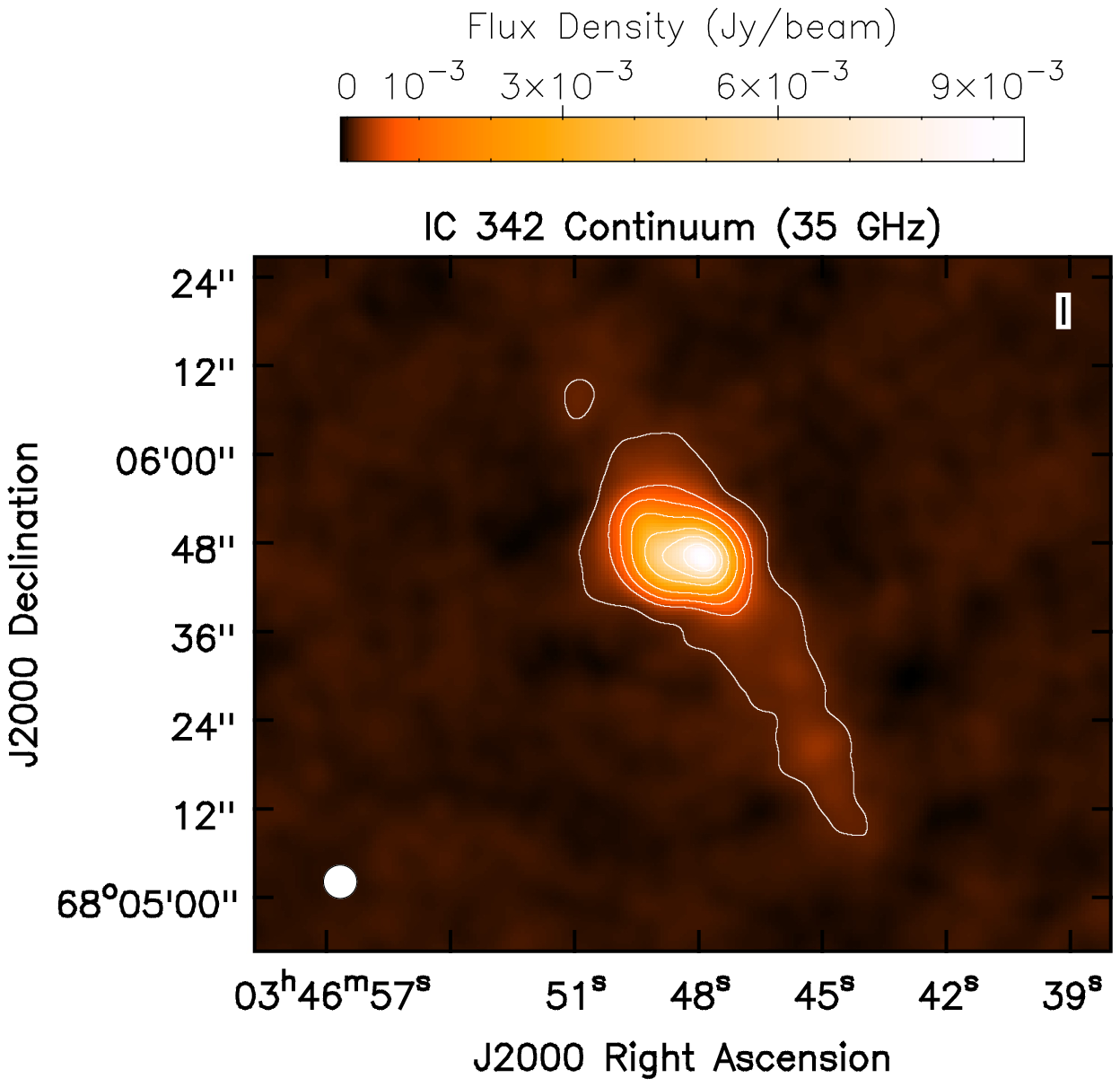} 
\includegraphics[angle=0,scale=0.8]{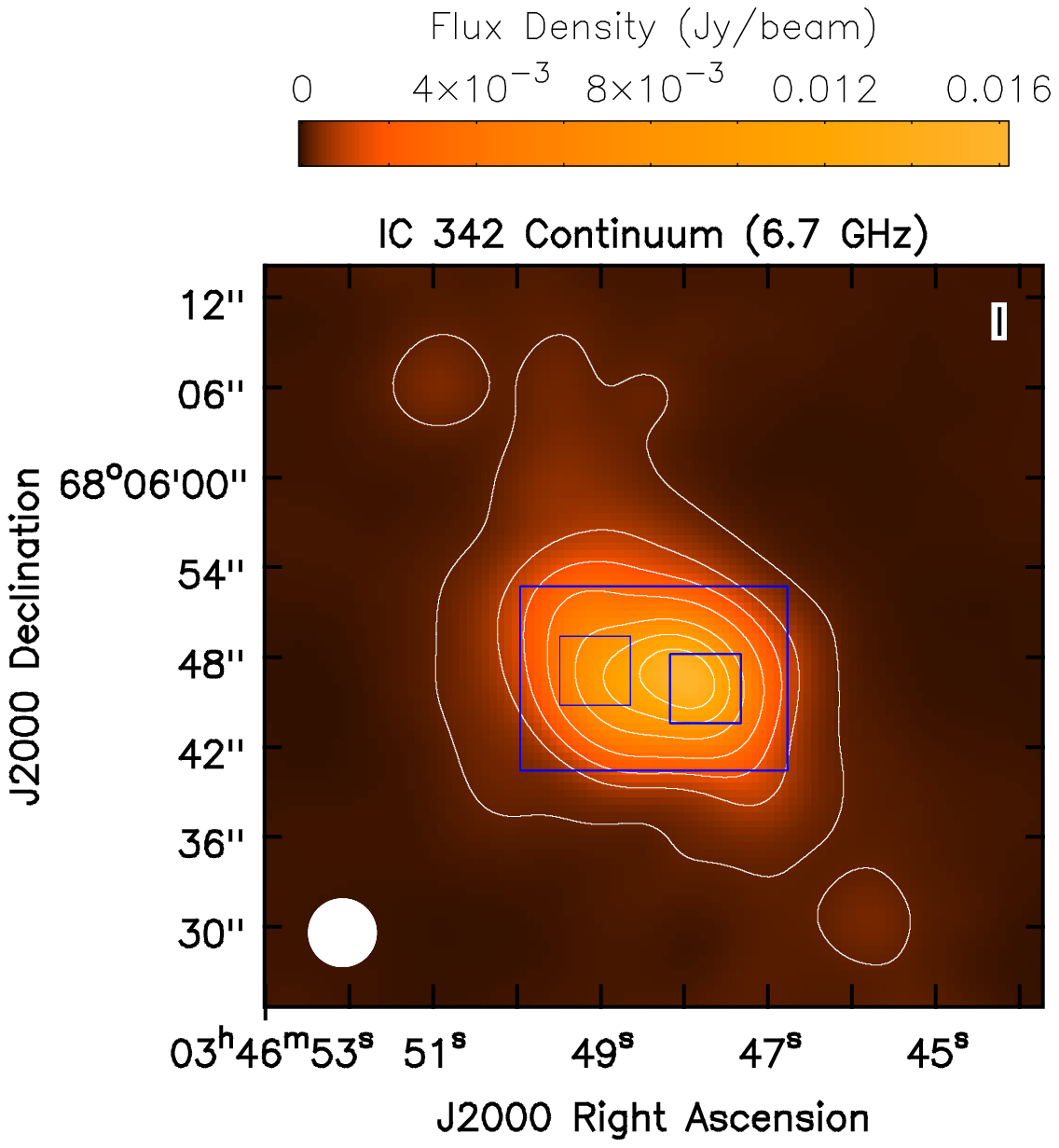} 
\caption{{\it Top:} JVLA continuum image of \ic{342} at 35\ghz.  The
  contour levels are 0.01, 0.05, 0.1, 0.2, 0.4, 0.6, 0.8 times the
  peak value of 9.45\mjyb.  The rms noise in the image is 13\microjyb.
  The synthesized beam size, shown in the bottom-left corner, is
  $4\arcsper5 \times 4\arcsper5$.  {\it Bottom:} JVLA continuum image
  of \ic{342} at 6.7\ghz.  The contour levels are 0.01, 0.05, 0.1,
  0.2, 0.4, 0.6, 0.8 times the peak value of 16.25\mjyb.  The rms noise
  in the image is 6.5\microjyb.  The synthesized beam size, shown in
  the bottom-left corner, is $4\arcsper5 \times 4\arcsper5$.  The blue
  rectangles correspond to the three regions used in the analysis.
  The ``whole region'' is the larger box.  The ``East'' and ``West''
  components correspond to the left and right boxes, respectively.}
\label{fig:cont}
\end{figure}

\begin{figure}
\includegraphics[angle=0,scale=0.45]{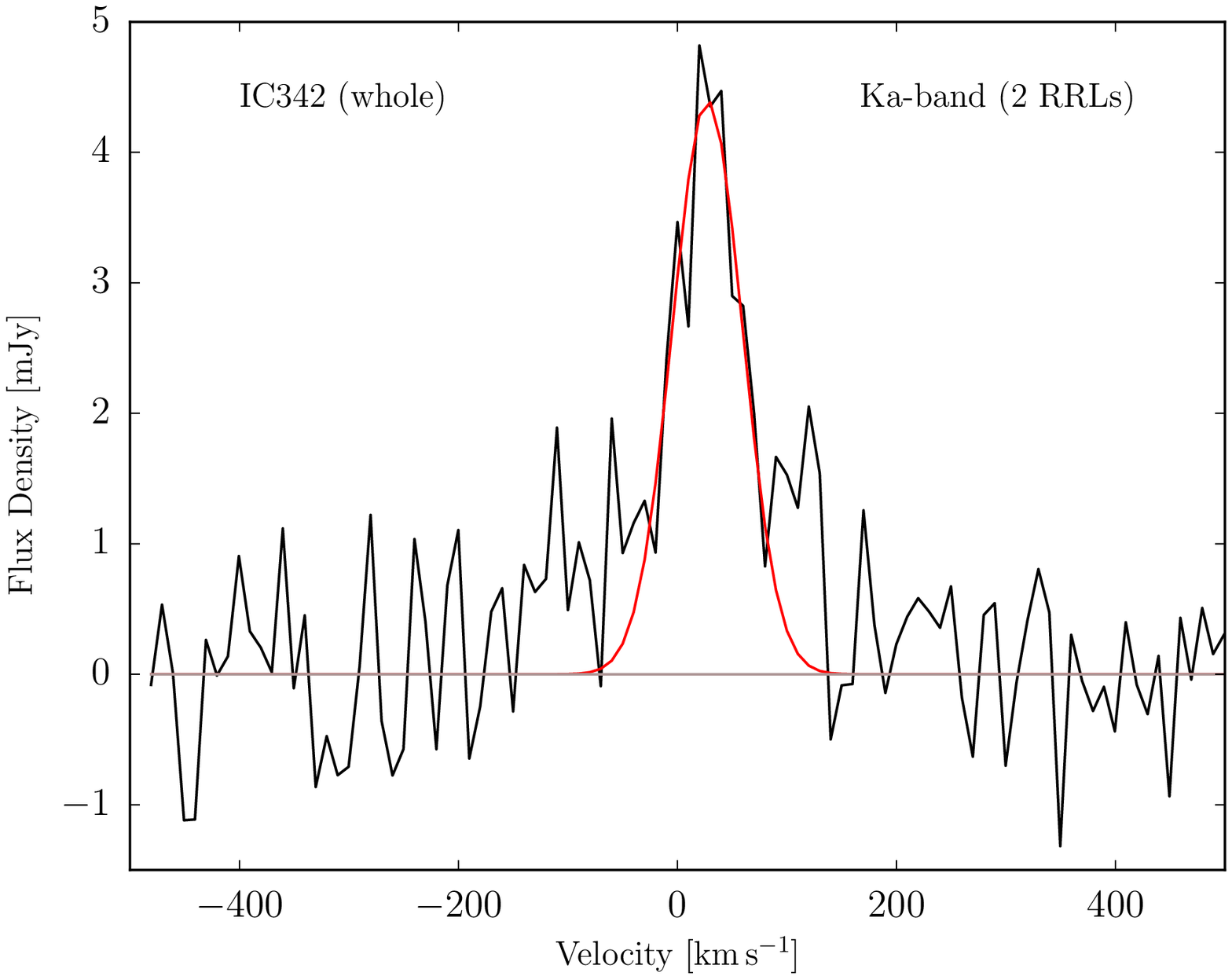} 
\includegraphics[angle=0,scale=0.45]{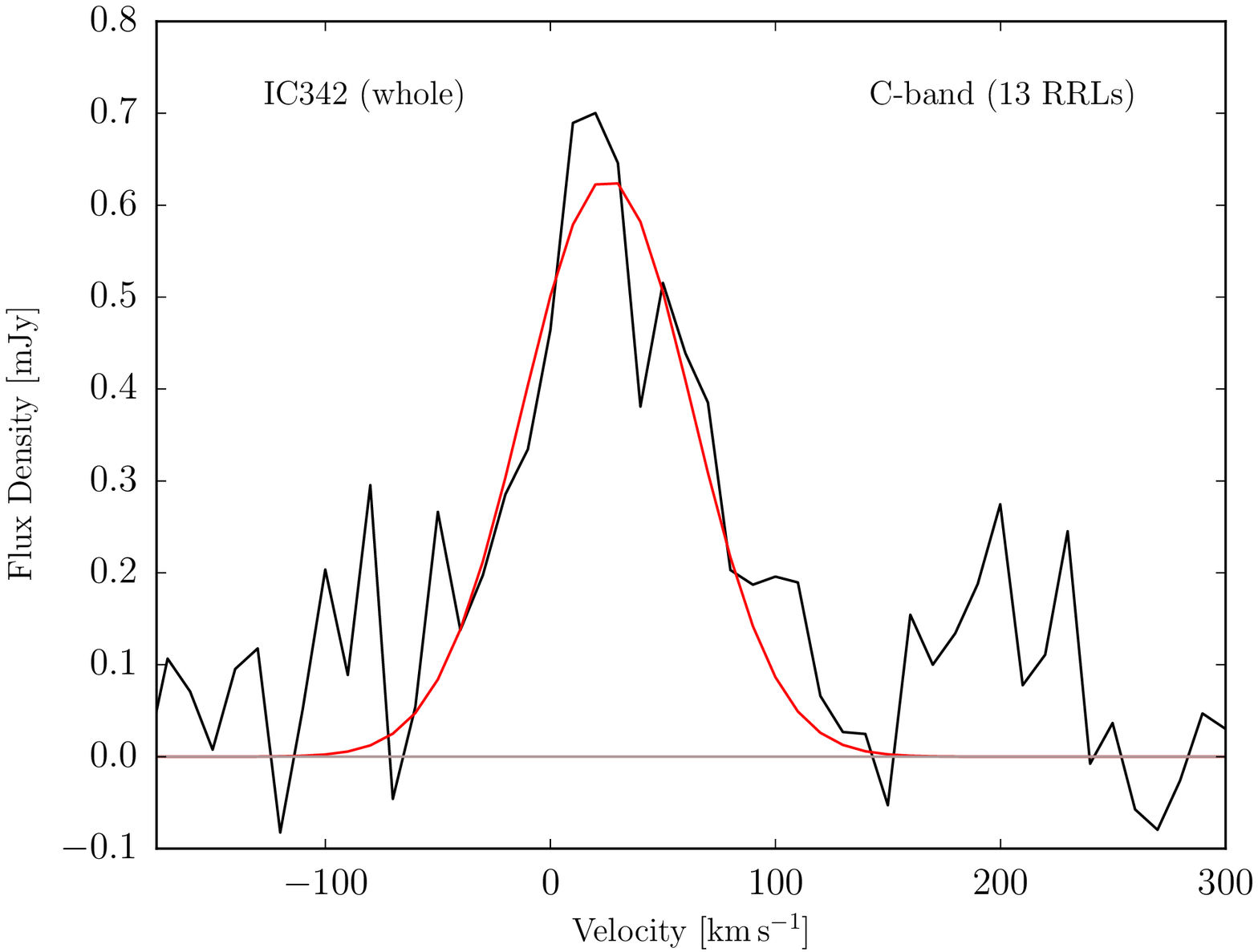} 
\includegraphics[angle=0,scale=0.45]{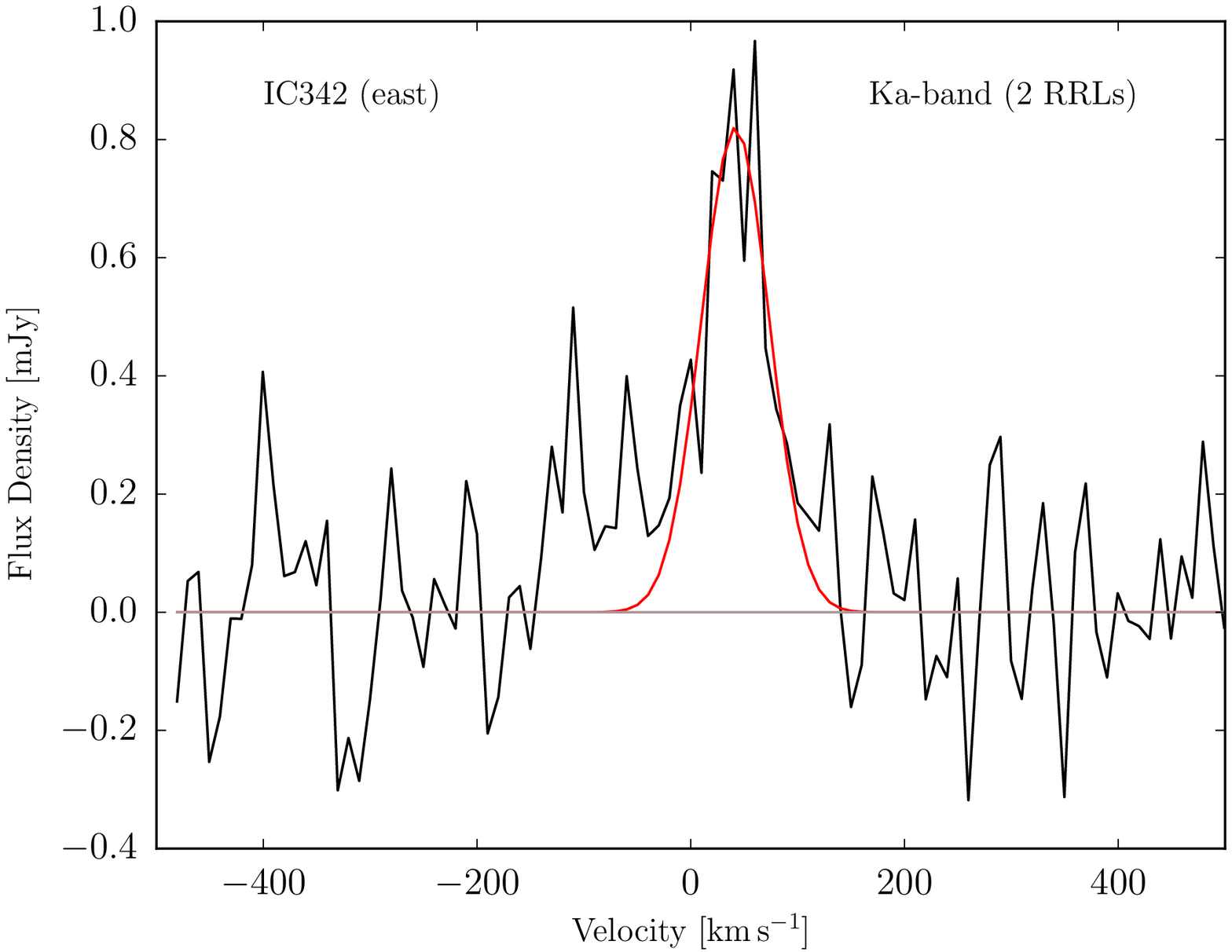} 
\includegraphics[angle=0,scale=0.45]{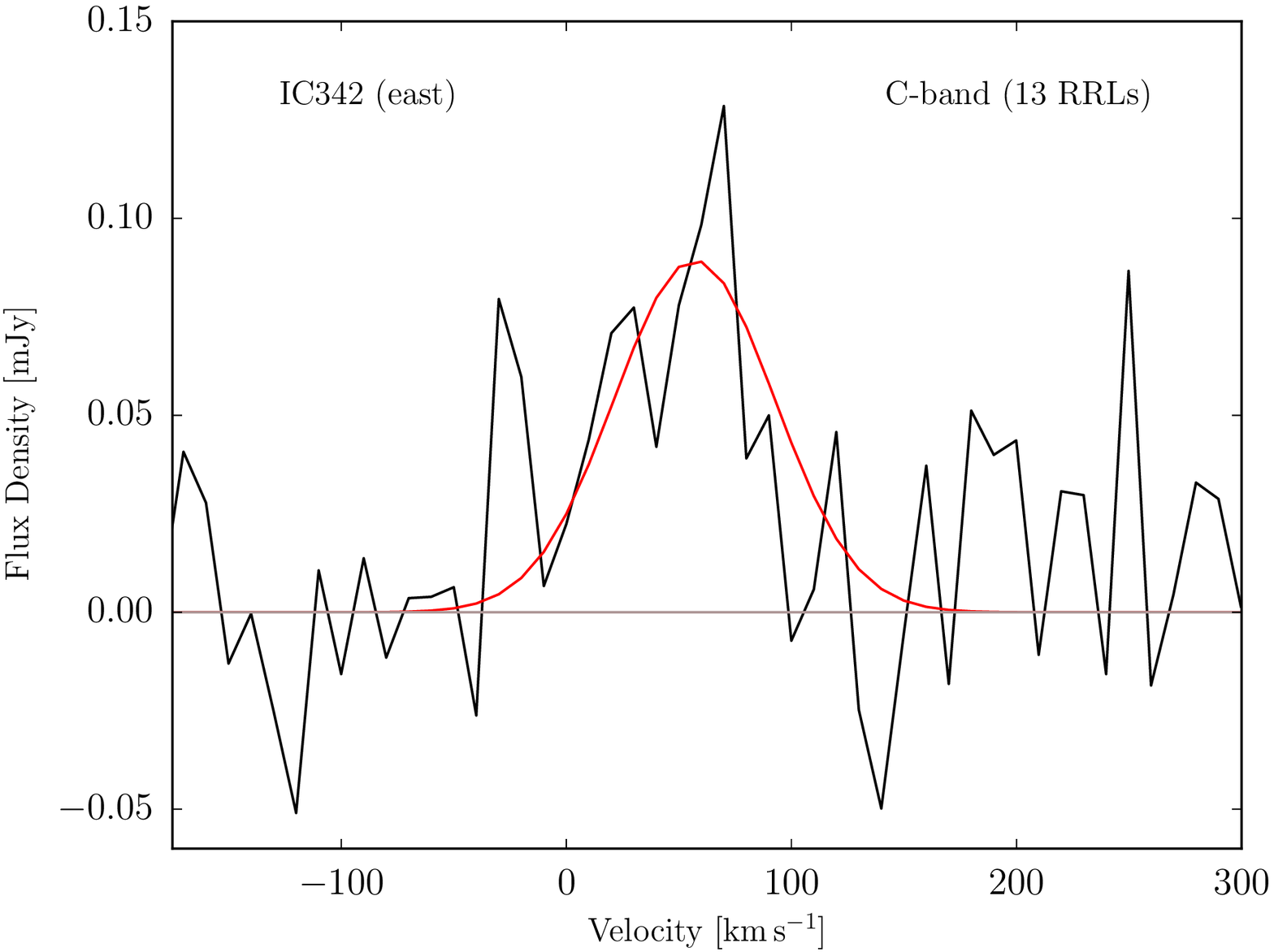} 
\includegraphics[angle=0,scale=0.45]{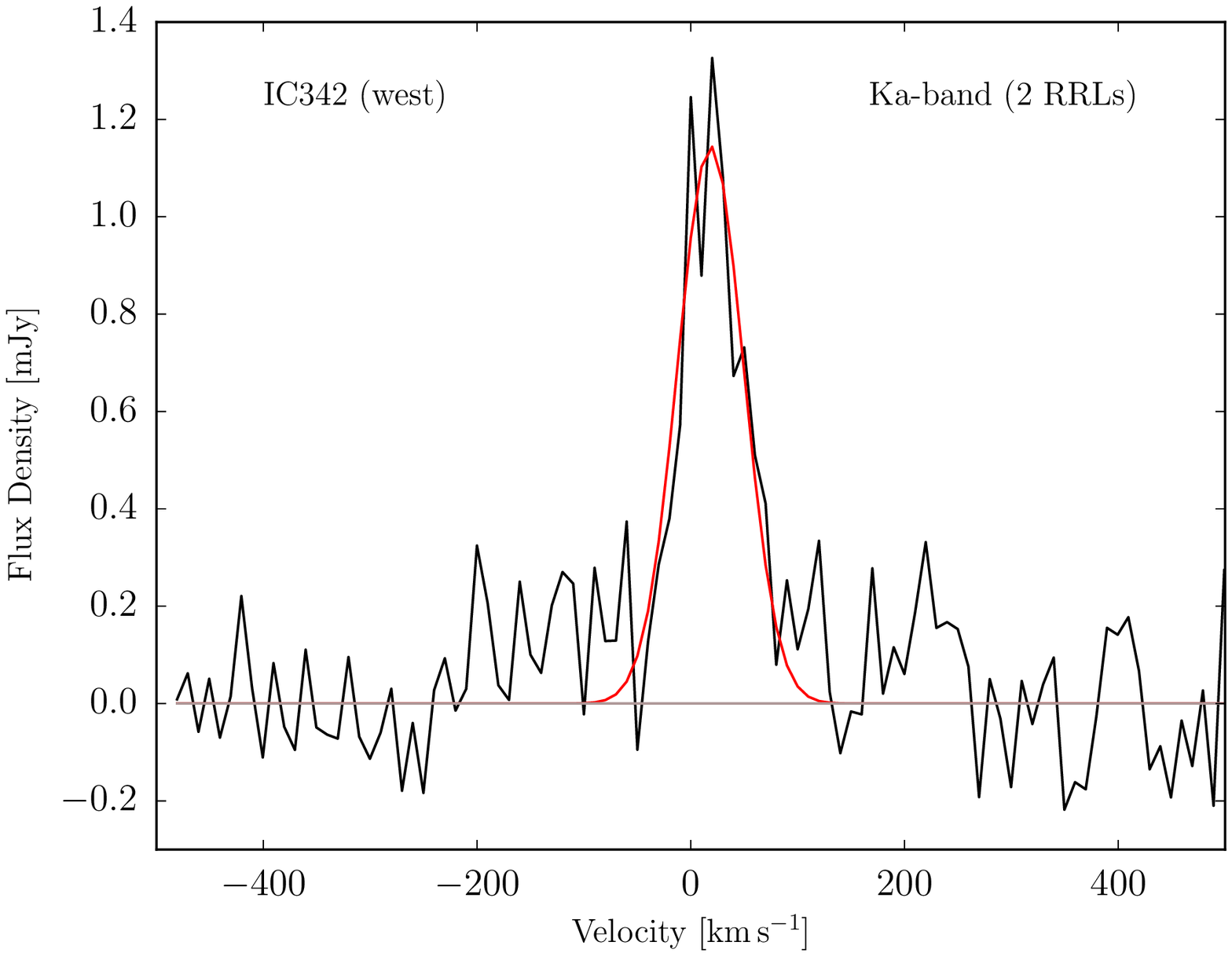} 
\includegraphics[angle=0,scale=0.45]{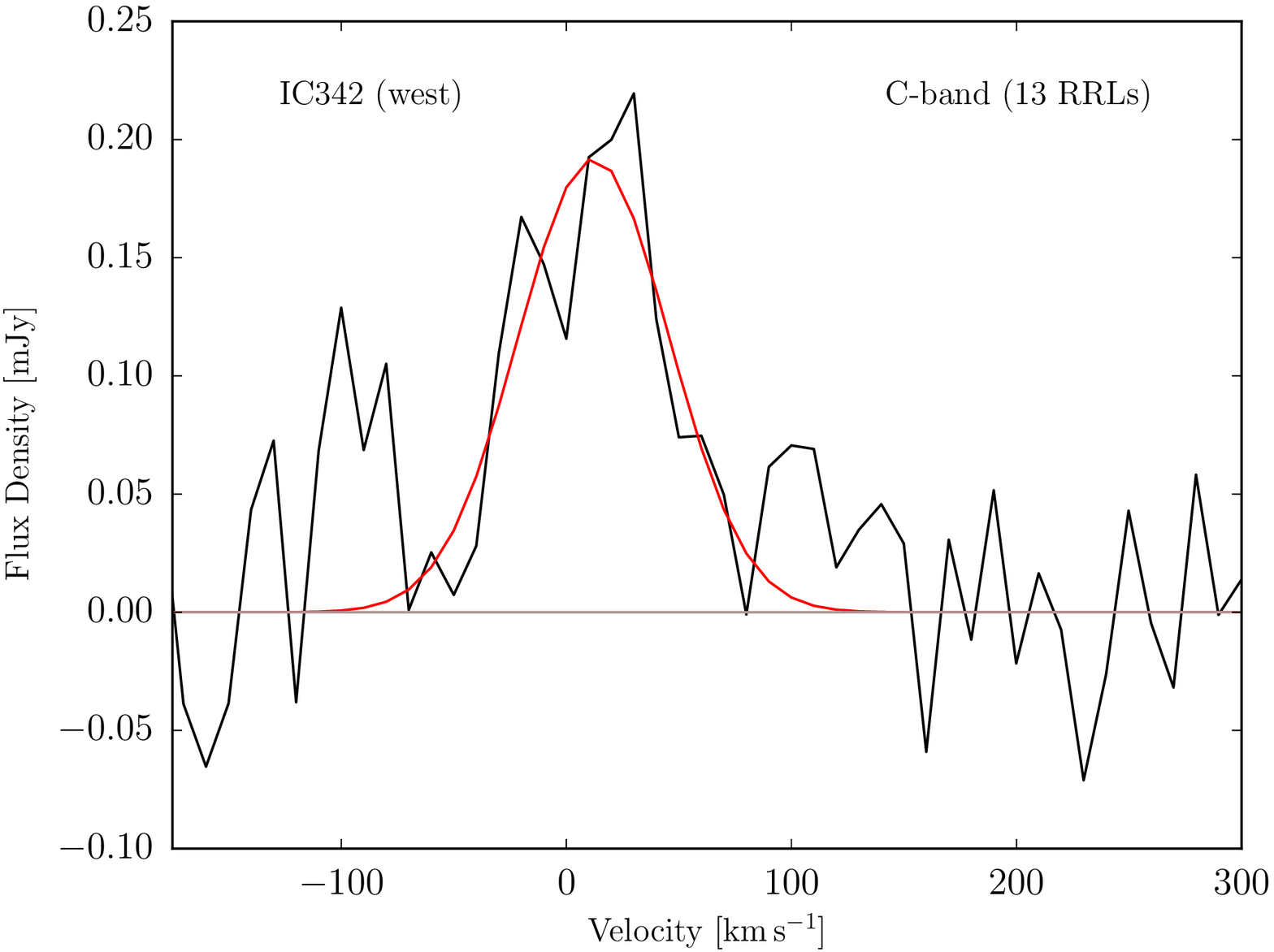} 
\caption{Integrated RRL spectra of the whole region (top), East
  component (middle), and West component (bottom).  The flux density
  integrated over each region is plotted as a function of the
  Barycentric velocity.  The red curve is a Gaussian fit to the data.
  {\it Left:} Integrated RRL spectrum at Ka-band
  (\hrrl{57}-\hrrl{58}).  {\it Right:} Integrated RRL spectrum at
  C-band (\hrrl{97}-\hrrl{101} and \hrrl{106}-\hrrl{113}).}
\label{fig:spectra}
\end{figure}

\begin{figure}
\includegraphics[angle=0,scale=0.85]{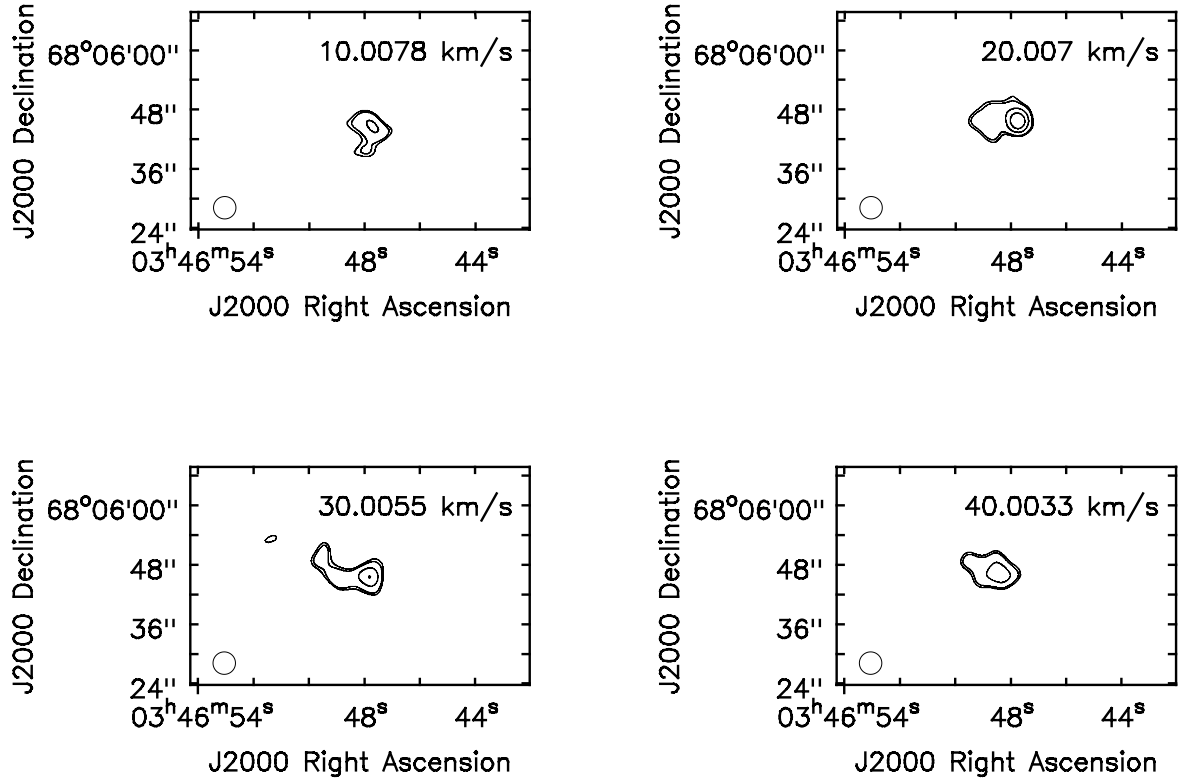} 
\caption{Ka-band RRL channel images.  The contours are 0.35, 0.4, 0.6,
  and 0.8 times the peak emission of 1.74\mjyb.  The lowest contour
  level is $3\,\sigma$, where $\sigma \sim 0.2$\mjyb\ is the rms noise
  in a single channel image.  The Barycentric velocity (optical
  definition) is shown in the upper right-hand corner of each image.
  The synthesized beam size ($4\arcsper5 \times 4\arcsper5$) is shown
  in the bottom left-hand corner of each image.}
\label{fig:channel_images}
\end{figure}

\begin{figure}
\includegraphics[angle=0,scale=0.5]{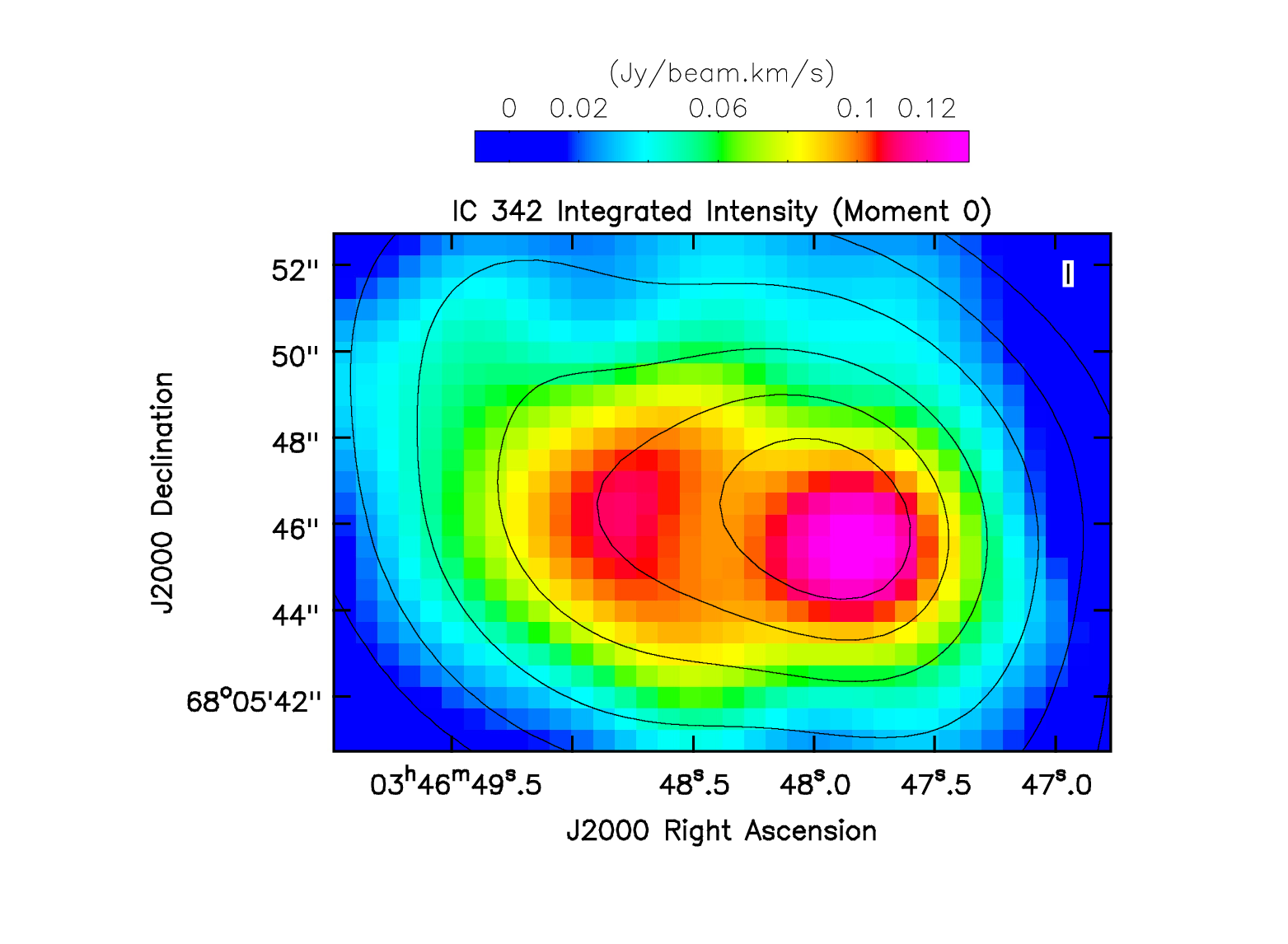} 
\caption{Ka-band RRL integrated intensity map (moment 0).  The
  synthesized beam size, not shown for clarity, is $4\arcsper5 \times
  4\arcsper5$.  The contours are the Ka-band continuum emission shown
  in Figure~\ref{fig:cont}.  The RRL emission is restricted to the
  nucleus and there is no significant RRL integrated intensity beyond
  the region shown.}
\label{fig:moment}
\end{figure}

The JVLA continuum images are shown in Figure~\ref{fig:cont} where we
only detect emission in the nuclear region of \ic{342}.  We select
three regions for analysis based on inspection of the spectral line
cubes (see below).  The large blue rectangle shown in
Figure~\ref{fig:cont} contains the ``whole region'' as defined by the
RRL emission.  The smaller blue rectangles correspond to different RRL
features within the channel maps and are denoted as the ``East
component'' and the ``West component''.  We sum emission over these
spatial regions for further analysis and modeling.  Integrated RRL
spectra are shown in Figure~\ref{fig:spectra} over the whole region,
the East component, and the West component.  We detect RRL emission at
each of the four frequencies listed above, but have stacked all RRLs
within each receiver band to increase the signal-to-noise ratio.  The
Ka-band RRLs are stacked using the same procedure as for the C-band
RRLs (see \S{\ref{obs:imaging}}).  This gives us two RRL data points
separated over a wide frequency range for modeling.  To our knowledge
this is the first detection of RRL emission in \ic{342}.  The RRL
profiles are best fit by a single Gaussian with a full-width
half-maximum (FWHM) line width $\Delta{V} = 70-90$\kms.  The East
component has a center velocity $V_{\rm bary} \sim 45$\kms,
significantly higher than the West component with $V_{\rm bary} \sim
15$\kms.  The velocity structure between the East and West components
is shown in channel images taken from the Ka-band RRL data cubes
(Figure~\ref{fig:channel_images}).  The extent of the RRL emission is
restricted to the central regions of the JVLA image (See
Figure~\ref{fig:moment}).

Tables \ref{tab:cont}-\ref{tab:line} summarize the radio continuum and
RRL results.  We list the total continuum flux density (thermal plus
non-thermal), integrated over the whole region, the East component,
and the West component.  We estimate a 10\% error in the total
continuum flux density.  The Gaussian fit parameters to the spectra in
Figure~\ref{fig:spectra} are listed in Table~\ref{tab:line} together with
the RRL flux density integrated over the line profile.

\begin{deluxetable}{lcrr}
\tablecaption{Radio Continuum Results\label{tab:cont}}
\tablehead{
\colhead{} & \multicolumn{3}{c}{\underline{~~~~~~~~~$S^{\rm tot}_{\rm C}$ (mJy)~~~~~~~~~}} \\
\colhead{Freq.} & \colhead{Whole} & \colhead{East} & \colhead{West} \\
\colhead{(GHz)} & \colhead{Region} & \colhead{Comp.} & \colhead{Comp.}
}
\startdata
5.0 & 61.1 & 10.1 & 13.5 \\
6.7 & 51.6 &  8.4 & 11.6 \\
34  & 29.0 &  4.0 &  6.9 \\
35  & 28.6 &  3.9 &  6.9 \\
\enddata
\end{deluxetable}

\begin{deluxetable}{lcccc}
\tablecaption{Radio Recombination Line Results\label{tab:line}}
\tablehead{
\colhead{} & \colhead{} & \multicolumn{3}{c}{\underline{~~~~~~~~~~~~~Gaussian Fit\tablenotemark{a}~~~~~~~~~~~~~~~~~~~~~~}} \\
\colhead{Freq.} & \colhead{$S_{\rm L}^{\rm int}$} &
\colhead{$S_{\rm L}^{\rm pk}$} & \colhead{$\Delta V$} & \colhead{$V_{\rm bary}$} \\ 
\colhead{(GHz)} & \colhead{(mJy$\,$\kms)} & \colhead{(mJy)} & \colhead{(\kms)} & \colhead{(\kms)} 
}
\startdata
\multicolumn{5}{c}{\underline{Whole Region}} \hfill \\
6   & $59.1 \pm 10.8$  & $0.629 \pm 0.054$ & $88.3 \pm 14.3$ & $25.3 \pm 4.4$ \\
35  & $352.1 \pm 47.6$ & $4.398 \pm 0.032$ & $75.3 \pm  8.5$ & $27.4 \pm 2.9$ \\
\multicolumn{5}{c}{\underline{East Component}} \hfill \\
6   & $8.0 \pm 3.0$    & $0.089 \pm 0.014$ & $83.9 \pm 28.7$ & $56.9 \pm 11.2$ \\
35  & $65.1 \pm 14.8$  & $0.820 \pm 0.102$ & $74.5 \pm 14.2$ & $41.7 \pm 4.8$ \\
\multicolumn{5}{c}{\underline{West Component}} \hfill \\
6   & $16.1 \pm 3.2$   & $0.192 \pm 0.020$ & $79.0 \pm 13.3$ & $12.1 \pm 4.9$ \\
35  & $88.6 \pm 16.6$  & $1.145 \pm 0.110$ & $72.7 \pm 11.7$ & $18.5 \pm 4.3$ \\
\enddata
\tablenotetext{a}{Listed are the peak intensity, $S_{\rm L}^{\rm pk}$,
  the full-width half-maximum (FWHM) line width, $\Delta V$, and the
  Barycentric velocity, $V_{\rm bary}$.}
\end{deluxetable}

\section{\hii\ Region Models}

We fit the JVLA data of the nucleus of \ic{342} with two different
models, assuming a distance of 3.28\Mpc\ \citep{saha02}.  Models based
on RRL data typically fall into types: a uniform slab of ionized gas
or a collection of compact \hii\ regions.  In many cases the latter
model is favored since the uniform slab model often produces an excess
of thermal continuum emission \citep{anantharamaiah00}.

The first model is very simple and assumes that the nuclear region
specified in Figure~\ref{fig:cont} is fully ionized with gas at a
constant electron temperature, \te, and electron density, \Ne.  \hii\
regions are primarily heated by the hydrogen-ionizing photons from
early-type stars and cooled by collisionally excited lines (e.g.,
[CII]158\microns).  For OB stars the electron temperature is set
primarily by the metallicity \citep{rubin85} with typical values near
$T_{\rm e} = 7500$\K\ in the central regions of the Milky Way
\citep{balser11}.  Following \citet{puxley91}, we assume that the RRL
emission is formed from only spontaneous emission and that the gas is
in LTE.  That is, there is no stimulated emission.  Here we call this
the Spontaneous Emission Model (SEM; see \S{\ref{sec:sem}} for
details).  Given a value of \Ne\ allows us to calculate the integrated
RRL flux and the thermal continuum flux density.  Assuming that the
non-thermal continuum flux density follows a power law ($S_{\rm nth}
\propto \nu^{\alpha}$), we use the observed total continuum flux
density to calculate the spectral index, $\alpha$.

The RRL data fit the model well for low electron densities $n_{\rm e}
= 25-70$\percc.  For spontaneous, optically thin emission in LTE the
integrated line flux is $\propto \nu_{\rm L}^2$
\citep[e.g.,][]{puxley97}, which is consistent with the our results.
Using the electron density that best fits the data we derive the
corresponding continuum flux density.  For each region the predicted
continuum flux density is larger than is observed, by as much as a
factor of 2.5, and therefore inconsistent with the data.  Regardless,
we use the RRL results to estimate the star formation rate for this
model.  In \S{\ref{sec:sfr}}, we derive the total number of
hydrogen-ionizing photons, $N_{\rm L}$, by balancing the ionization
rate versus the recombination rate, assuming a spherical \hii\ region
with diameter equal to the line emitting region size.  For the whole
region the line emitting size is $L = 132$\pc.  For $n_{\rm e} \sim
25$\percc\ we get $N_{\rm L} = 7.2 \times 10^{51}\,{\rm s}^{-1}$.
This yields a star formation rate of SFR $= 0.053$\sfr\
\citep{murphy11}.

Next, we assume that the nuclear region specified in
Figure~\ref{fig:cont} contains many compact \hii\ regions that
uniformly fill this volume.  Following \citet{anantharamaiah93}, we
assume each \hii\ region is identical with a constant electron
temperature, \te, electron density, \Ne, and linear size, $\ell$.  We
call this model the multiple \hii\ regions model (MHM; see
\S{\ref{sec:mhm}} for details).  The linear size of the line emitting
region, $L$, is set by the solid angle of the different regions
specified in Figure~\ref{fig:cont}.  This model includes non-LTE
effects with pressure broadening from electron impacts.  There are
three free parameters: \te, \Ne, and $\ell$.  As with the SEM we
assume an electron temperature which leaves two data points and two
unknowns to constrain the model.  We explore the effects of electron
temperature on the model by running the grid of (\Ne, $\ell$) for
several different fixed \te\ values.  The total number of \hii\
regions, $N_{\rm HII}$, is set by the ratio of the observed integrated
RRL flux to the flux of one \hii\ region.

The following constraints are used to reject any given model based on
physical plausibility \citep[see][]{anantharamaiah93}.

\begin{enumerate}

\item {\it Filling factor:} The \hii\ regions should be confined to
  the line emitting region and therefore the filling factor, $f =
  N_{\rm HII}\, \ell^{3}/L^{3}$, should be less than unity.

\item {\it Peak RRL Intensity:} The peak RRL intensity of a single
  \hii\ region in the model should be less than the observed peak RRL
  intensity which covers the entire volume.

\item {\it Minimum Number of \hii\ Regions:} The RRL width from a
  single \hii\ region in the model is much less than the observed RRL
  width due to the velocity motions of the compact \hii\ regions
  within the line emitting region.  Therefore the number of \hii\
  regions derived from the model, $N_{\rm HII}$, should exceed the
  minimum number given by $N_{\rm min} = (\Delta{V_{\rm
      obs}}\Omega_{\rm L}) / (\Delta{V_{\rm HII}}\Omega_{\rm B})$,
  where $\Omega_{\rm B}$ is the beam solid angle.

\item {\it Maximum Number of \hii\ Regions:} The volume of the line
  emitting region should be able to contain the number of \hii\
  regions constrained by the model.  Therefore the total volume of the
  $N_{\rm HII}$ \hii\ regions should be smaller than the volume of the
  line emitting region.  Assuming spherical regions this implies that
  $N_{\rm HII}$ should be less than $N_{\rm max} = (L/\ell)^{3}/N_{\rm
    HII}$.

\item {\it Thermal Continuum Flux Density:} The predicted
  thermal continuum flux density should be less than the observed
  (total) flux density.

\item {\it Spectral Index:} The measured
  non-thermal flux density should not be too steep (i.e., $\alpha >
  -1.5$; see \citet{lisenfeld00}).

\end{enumerate}

\begin{figure}
\includegraphics[angle=0,scale=0.31]{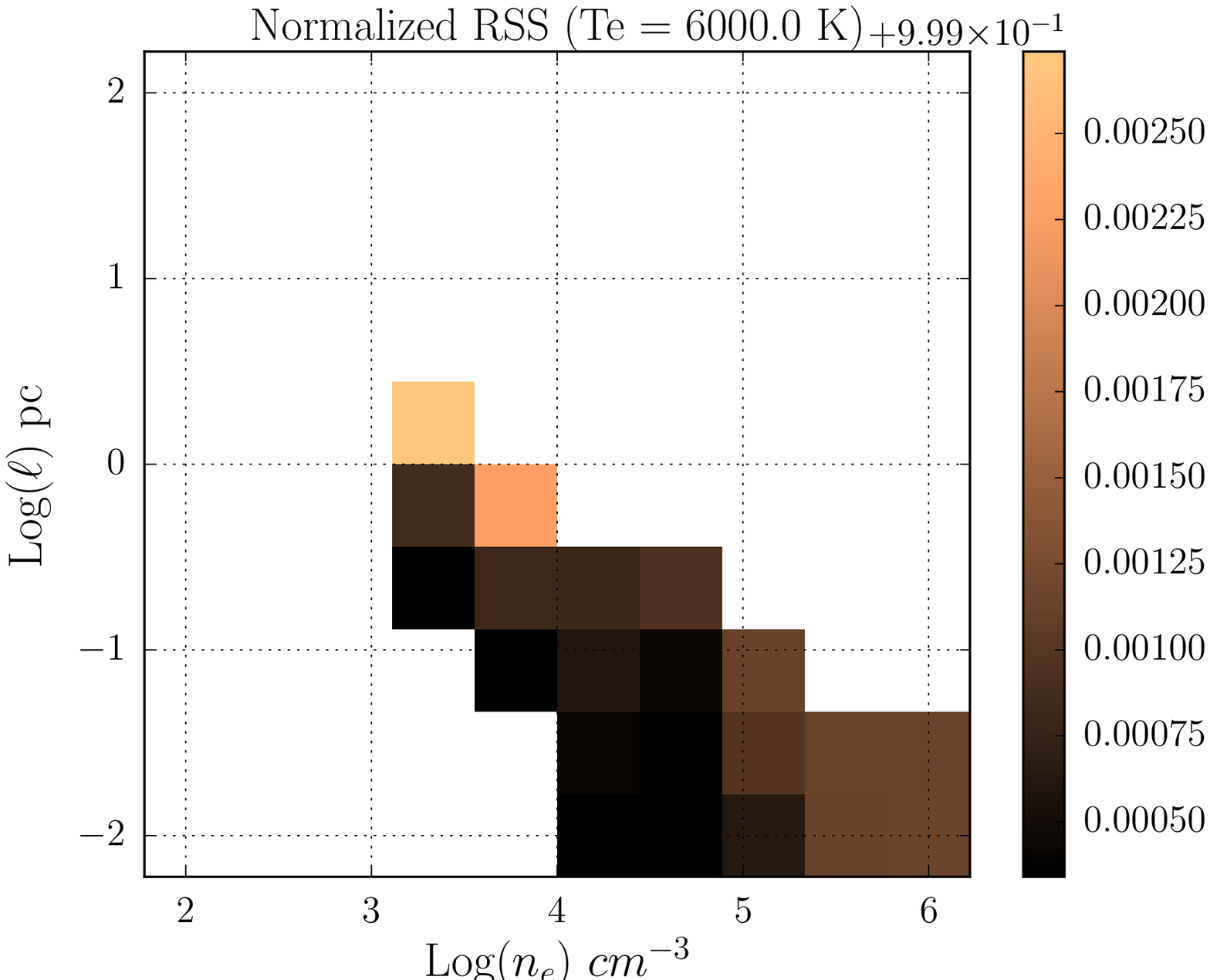} 
\includegraphics[angle=0,scale=0.31]{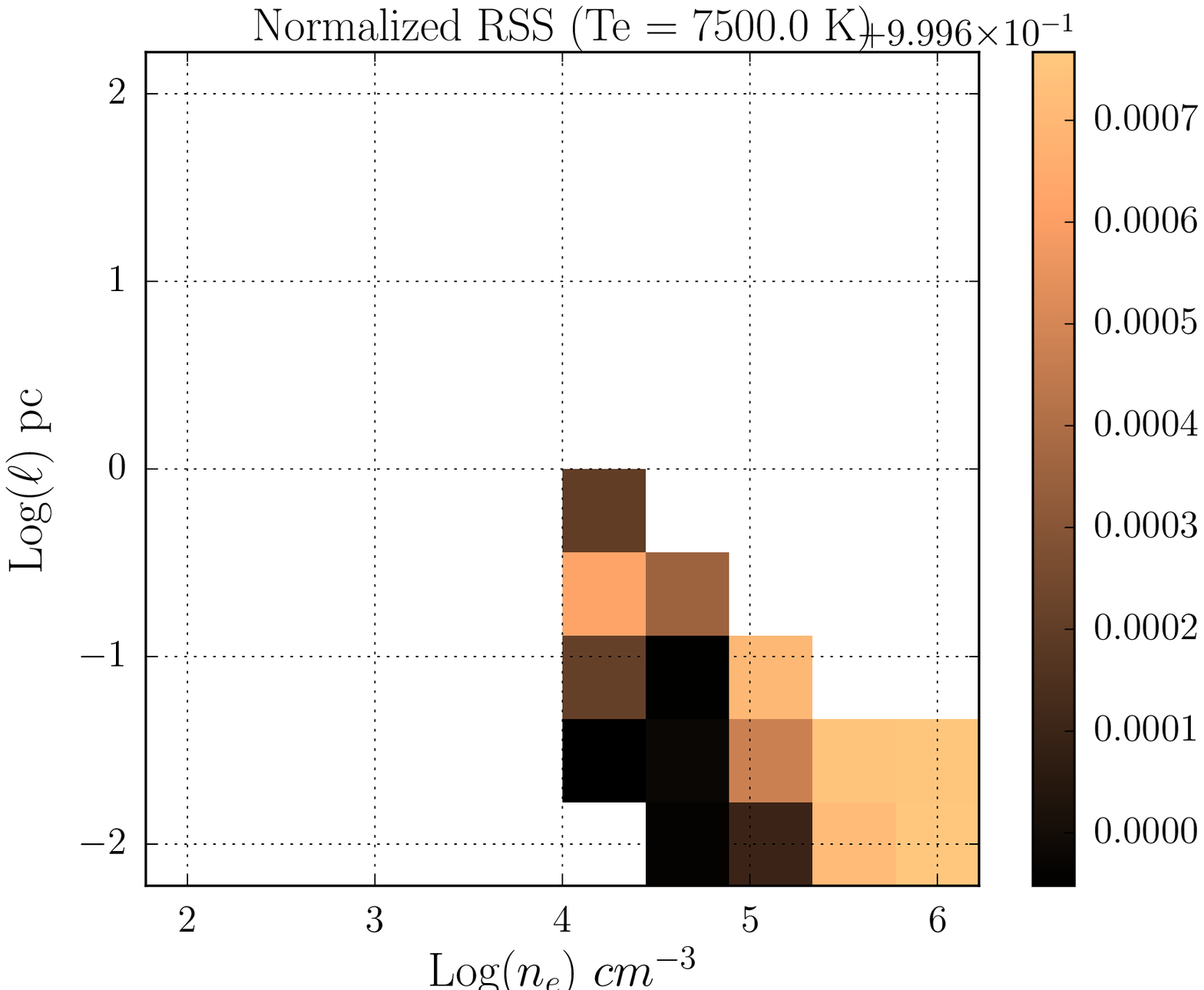} 
\includegraphics[angle=0,scale=0.31]{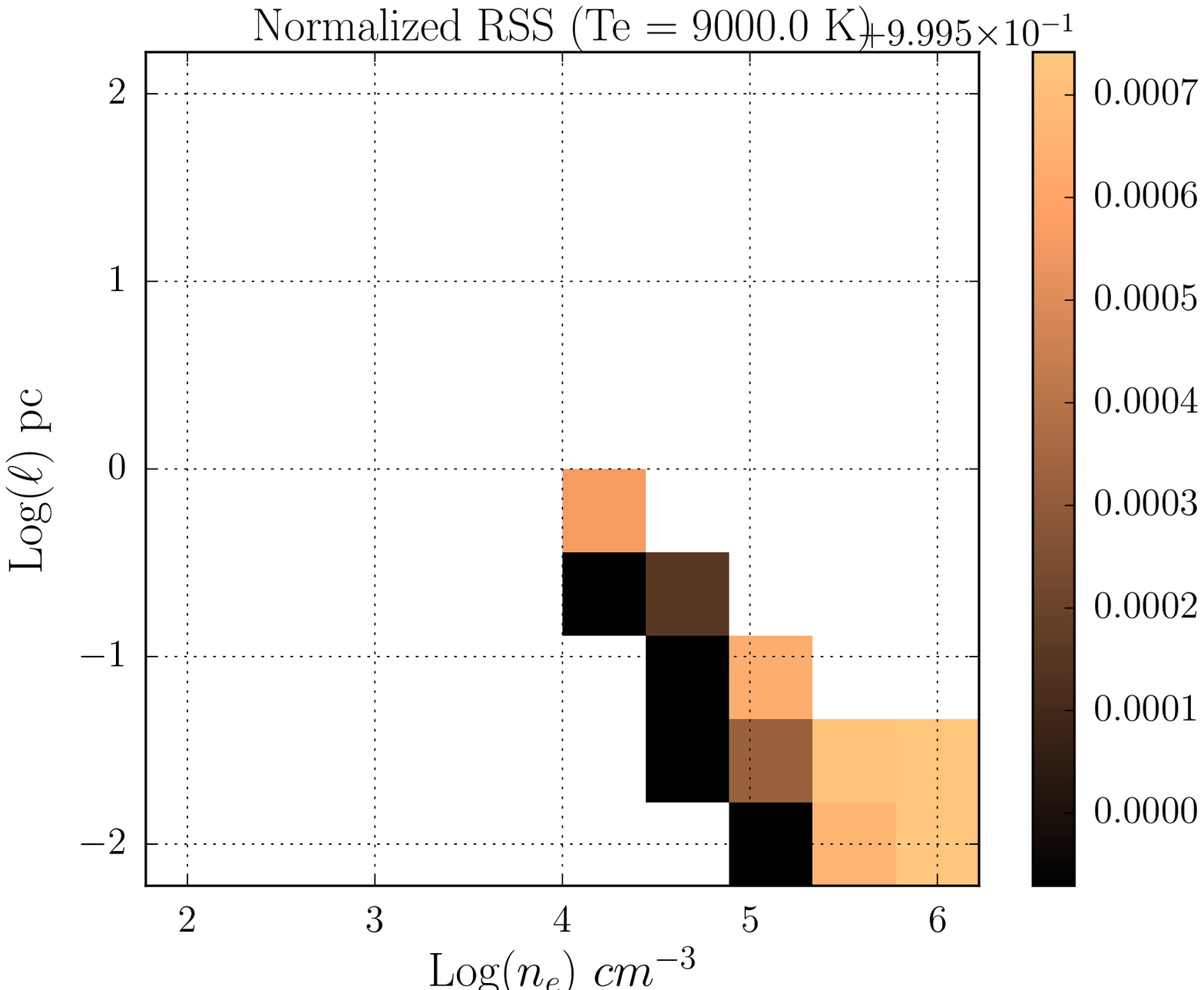} 
\includegraphics[angle=0,scale=0.3]{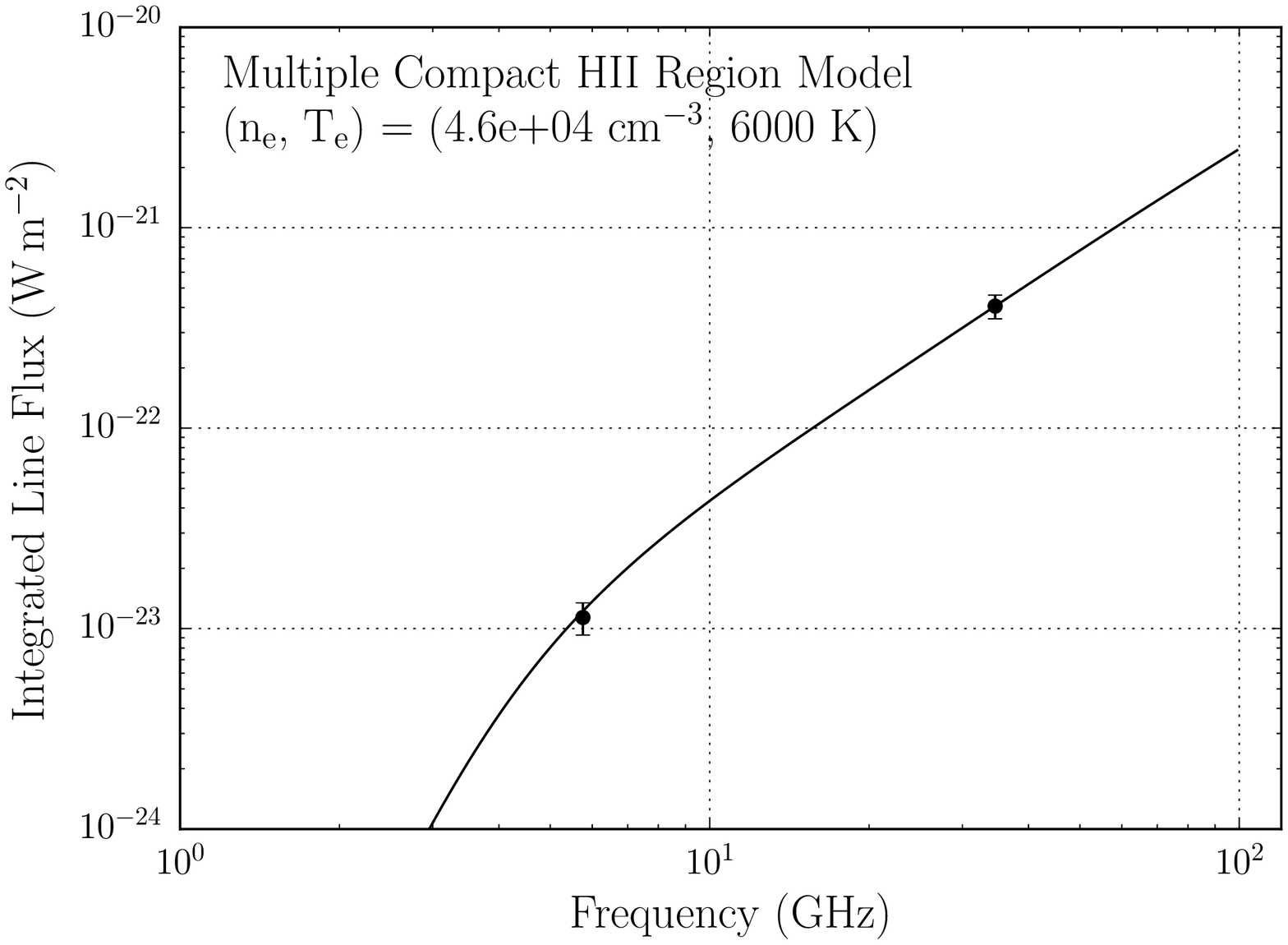} 
\includegraphics[angle=0,scale=0.3]{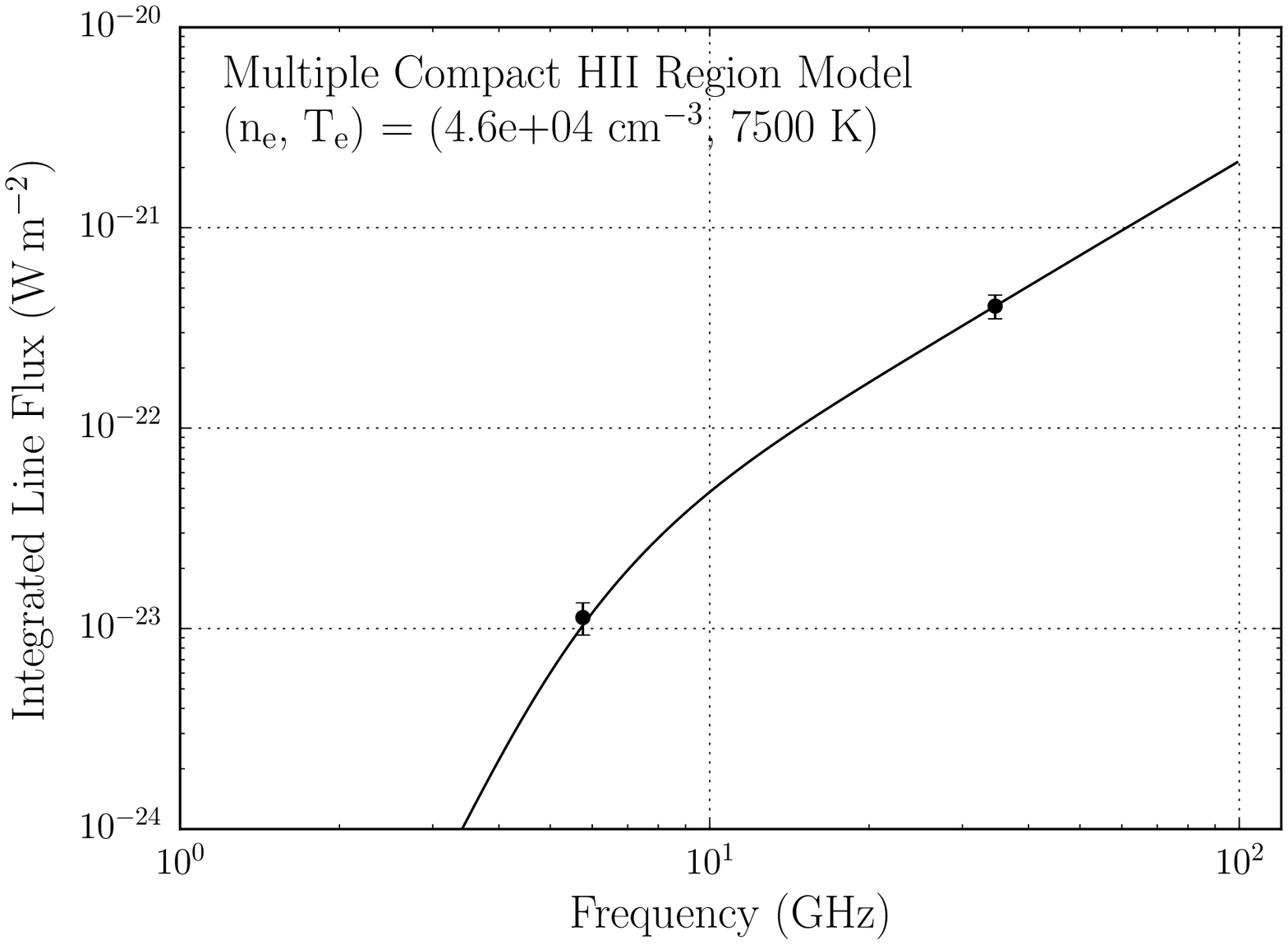} 
\includegraphics[angle=0,scale=0.3]{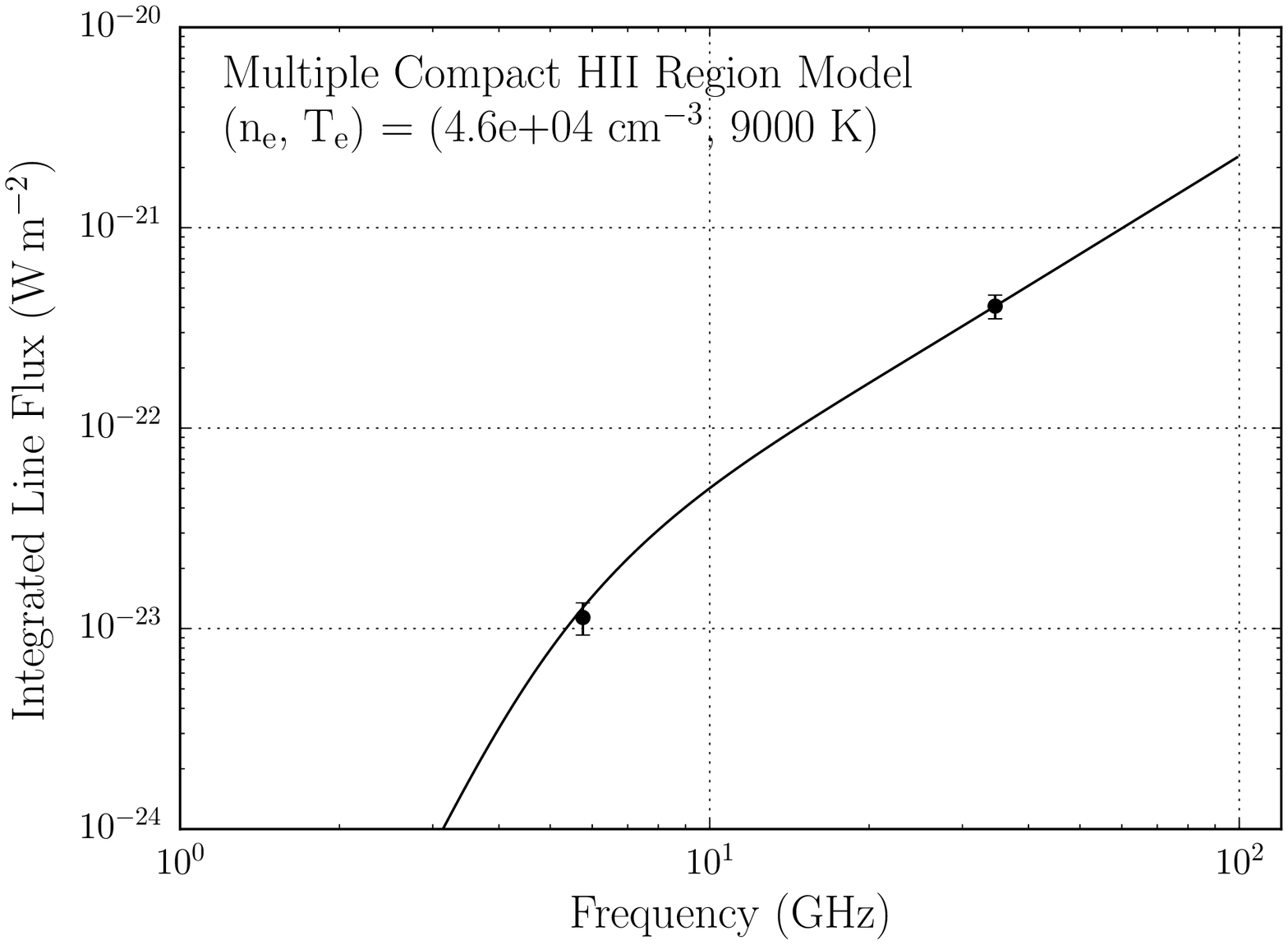} 
\includegraphics[angle=0,scale=0.3]{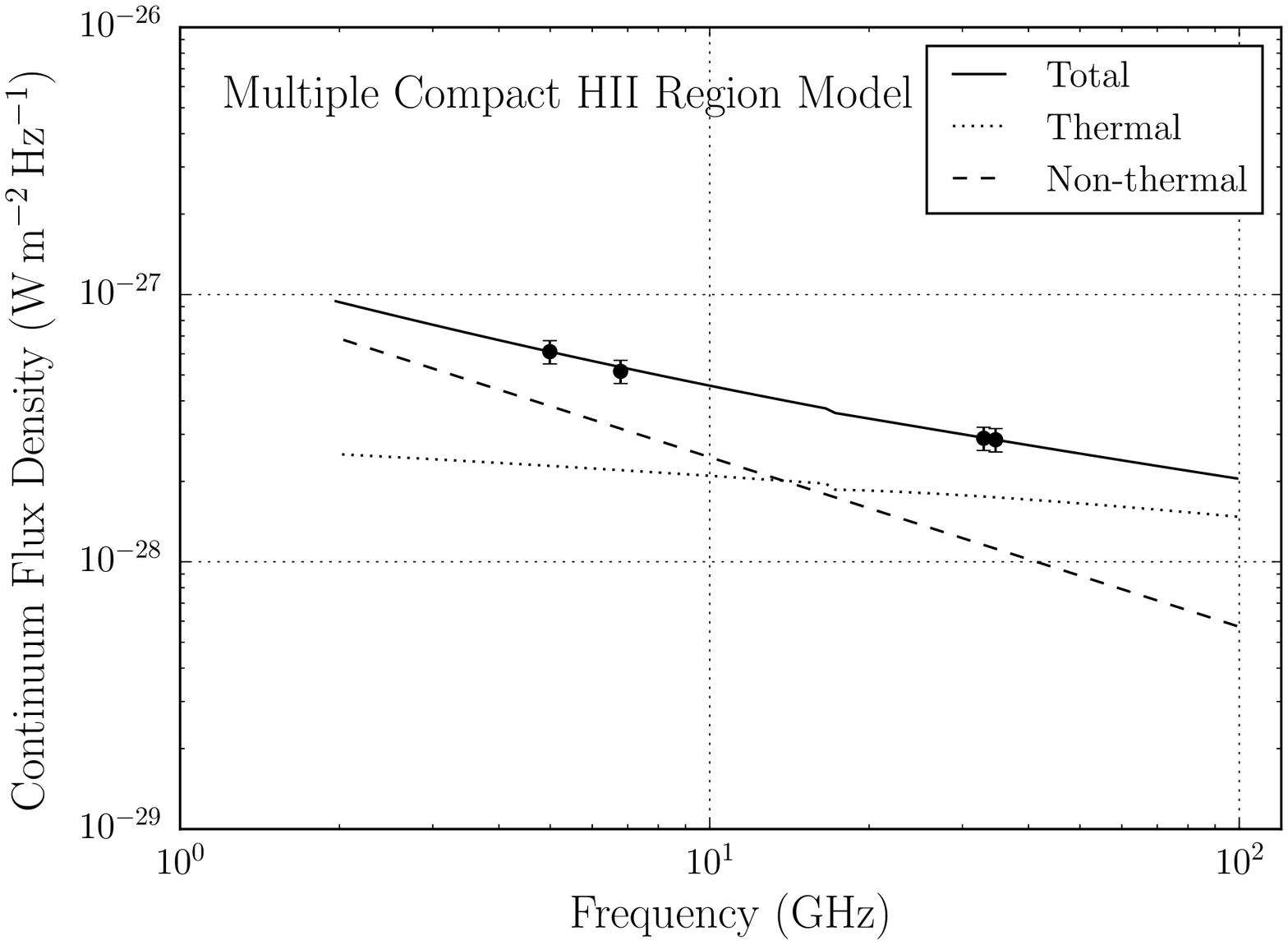} 
\includegraphics[angle=0,scale=0.3]{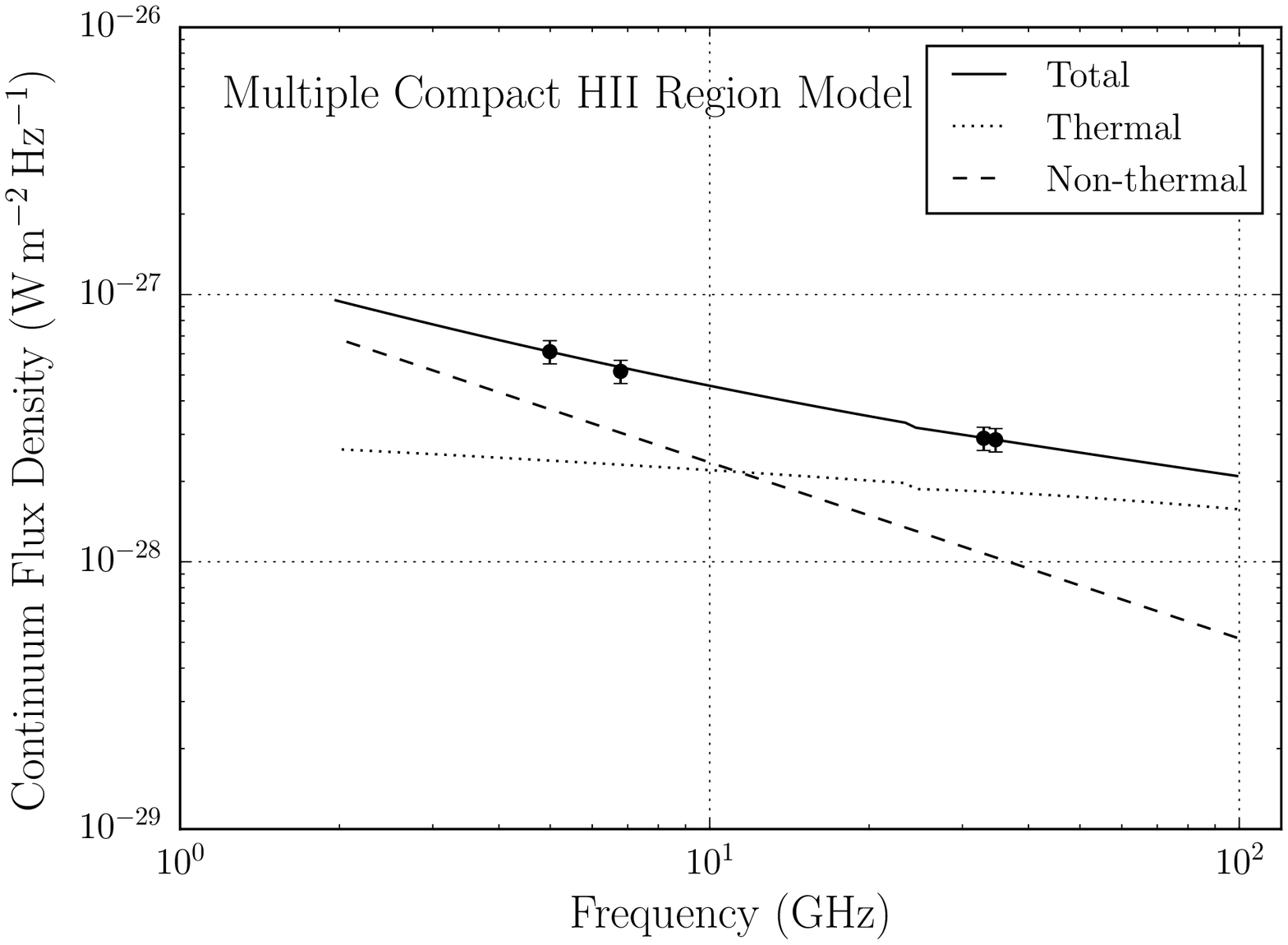} 
\includegraphics[angle=0,scale=0.3]{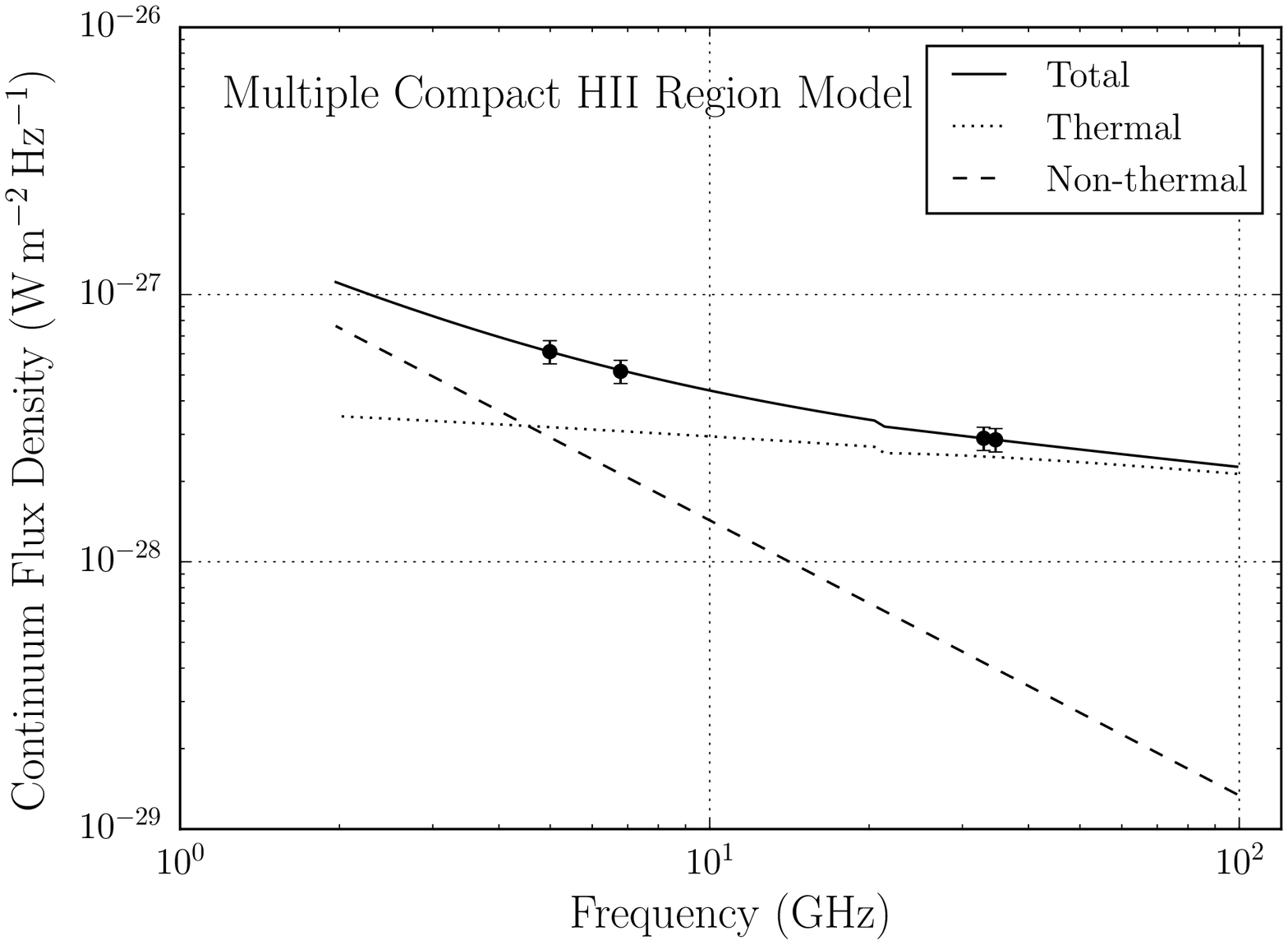} 
\caption{Results for the multiple \hii\ region model (MHM) for the
  whole region.  From left to right we explore different electron
  temperatures: $T_{\rm e} = 6000$\K, 7500\K, and 9000\K.  {\it Top:}
  Normalized root sum squared (RSS) values for the ($n_{\rm e}, \ell$)
  grid of models.  The blank areas are models that were not valid (see
  text). {\it Middle:} Integrated RRL flux as a function of frequency.
  The line corresponds to the best model fit and the points are JVLA
  data.  {\it Bottom:} Continuum flux density as a function of
  frequency.  The lines correspond to the thermal, non-thermal and
  total continuum flux density and the points are JVLA data.}
\label{fig:mhm_wholeRegion}
\end{figure}

A grid of models is run with $n_{\rm e} = 100-10^{6}$\percc\ and $\ell
= 0.01-100$\pc.  The results are shown in
Figure~\ref{fig:mhm_wholeRegion} for the whole region.  The best fit
model is determined by minimizing the root sum square (RSS) between
the data and model.\footnote{We are aware that minimizing the RSS may
  not be the best method to determine the ``solution'', but given the
  number of data points available we cannot carefully constrain these
  models and only seek a rough estimate of the star formation
  properties (see \citet{babu06} for a discussion of goodness of
  fit).}  Most models fail to pass our constraints.  The best models
lie along a diagonal line going from larger, diffuse \hii\ regions to
smaller, compact \hii\ regions.  Models above this line fail because
the peak line flux for a single \hii\ region is larger than the
observed peak line flux, whereas models below this line fail because
the number of \hii\ regions exceeds the maximum.  We use the Ka-band
integrated RRL flux to set the number of \hii\ regions and thus the
model goes through this data point. The C-band data may be optically
thick and therefore not detect all of the emission.

Table~\ref{tab:mhm} summarizes the results for the whole region, the
East component, and the West component.  The best models consist of
many compact ($\ell \sim 0.05$\pc), dense ($n_{\rm e} \sim 5 \times
10^{4}$\percc), \hii\ regions.  The total number of \hii\ regions is
often large ($N_{\rm HII} > 1000$), but models with fewer \hii\
regions (e.g., hundreds) and larger sizes (e.g., $\ell \sim 0.1$) also
fit the data well.  Regardless, the other properties (e.g., \Ne) are
not very sensitive to the \hii\ region size.  The electron temperature
does not have a strong effect on the results except for the number of
\hii\ regions and the non-thermal spectral index.  We expect the
electron temperature to be regulated primarily by the metallicity.
\citet{pilyugin14} measure 12 + log(O/H) = 8.8 in the central regions
of \ic{342} which corresponds to $T_{\rm e} = 6800$\K\
\citep[See][]{shaver83}.  But clearly we do not have sufficient data
to constrain \te\ (see Figure~\ref{fig:mhm_wholeRegion}).  Assuming
the continuum emission at 35\ghz\ is all thermal emission and we are
in LTE, the line-to-continuum ratio yields $T_{\rm e} = 7150$\K\
\citep{balser11}, consistent with expectations.  In all models the
non-thermal emission dominates at 6.7\ghz, whereas the thermal
emission dominates at 35\ghz.  The continuum data is well fit by
assuming the non-thermal emission can be characterized by a power law
where for most models $\alpha \sim -0.8$.

\begin{deluxetable}{lccc}
\tablecaption{Multiple \hii\ Region Model (MHM) Results\label{tab:mhm}}
\tablehead{
\colhead{} & \multicolumn{3}{c}{\underline{~~~~~Electron Temperature ($T_{\rm e}$)~~~~~}} \\
\colhead{Parameter} & \colhead{6000\K} & 
\colhead{7500\K} & \colhead{9000\K}
}
\startdata
\multicolumn{4}{c}{\underline{Whole Region}} \hfill \\
Electron Density ($n_{\rm e}$) \percc  & \nexpo{4.6}{4}   & \nexpo{4.6}{4}  & \nexpo{4.6}{4} \\
\hii\ Region Size ($\ell$) pc          & 0.028            & 0.077           & 0.077 \\
Number of \hii\ Region ($N_{\rm HII}$) & 66881            & 3494            & 4950 \\
Non-thermal Spectral Index ($\alpha$)  & $-0.63$          & $-0.66$         &  $-1.03$ \\
H-ionizing rate ($N_{\rm L}$) s$^{-1}$ & \nexpo{1.9}{52}  & \nexpo{1.8}{52} & \nexpo{2.2}{52} \\
Star Formation Rate (SFR) M$_{\odot}\,$yr$^{-1}$ & 0.14   & 0.13            & 0.16 \\
\multicolumn{4}{c}{\underline{East Component}} \hfill \\
Electron Density ($n_{\rm e}$) \percc  & \nexpo{4.6}{4}   & \nexpo{4.6}{4}  & \nexpo{1.3}{5} \\
\hii\ Region Size ($\ell$) pc          &  0.077           & 0.077           &  0.077 \\
Number of \hii\ Region ($N_{\rm HII}$) &  408             & 646             &  57 \\
Non-thermal Spectral Index ($\alpha$)  & $-0.75$          & $-1.2$          &  $-0.75$ \\
H-ionizing rate ($N_{\rm L}$) s$^{-1}$ & \nexpo{2.5}{51}  & \nexpo{3.3}{51} & \nexpo{1.9}{51} \\
Star Formation Rate (SFR) M$_{\odot}\,$yr$^{-1}$ & 0.018  & 0.024           & 0.014 \\
\multicolumn{4}{c}{\underline{West Component}} \hfill \\
Electron Density ($n_{\rm e}$) \percc  & \nexpo{4.6}{4}   & \nexpo{4.6}{4}  & \nexpo{4.6}{4} \\
\hii\ Region Size ($\ell$) pc          & 0.028            & 0.028           &  0.077 \\
Number of \hii\ Region ($N_{\rm HII}$) & 16823            & 24646           &  1245 \\
Non-thermal Spectral Index ($\alpha$)  & $-0.57$          & $-0.95$         &  $-1.03$ \\
H-ionizing rate ($N_{\rm L}$) s$^{-1}$ & \nexpo{4.7}{51}  & \nexpo{5.8}{51} & \nexpo{5.4}{51} \\
Star Formation Rate (SFR) M$_{\odot}\,$yr$^{-1}$ & 0.034  & 0.042           & 0.040 \\
\enddata
\tablecomments{These models are based on a distance to \ic{342} of 3.28\Mpc\ \citep{saha02}.} 
\end{deluxetable}

\newpage

\section{Discussion}

\ic{342} is a face-on, late-type spiral galaxy located in the
\ic{342}/Maffei Group of galaxies at a distance of 3.28\Mpc\
\citep{saha02}.  Because the disk is perpendicular to our
line-of-sight there is a clear view of the nuclear region which has
many similarities to the Galactic Center \citep{meier14}.  Both have a
central molecular zone, with a similar size and mass, that surrounds a
circumnuclear disk.  At the center is a star cluster producing a
similar number of hydrogen ionizing photons per second.  Therefore,
the nucleus of \ic{342} provides is a nice analog for comparison with
the Galactic Center.

\begin{figure}
\includegraphics[angle=0,scale=0.8]{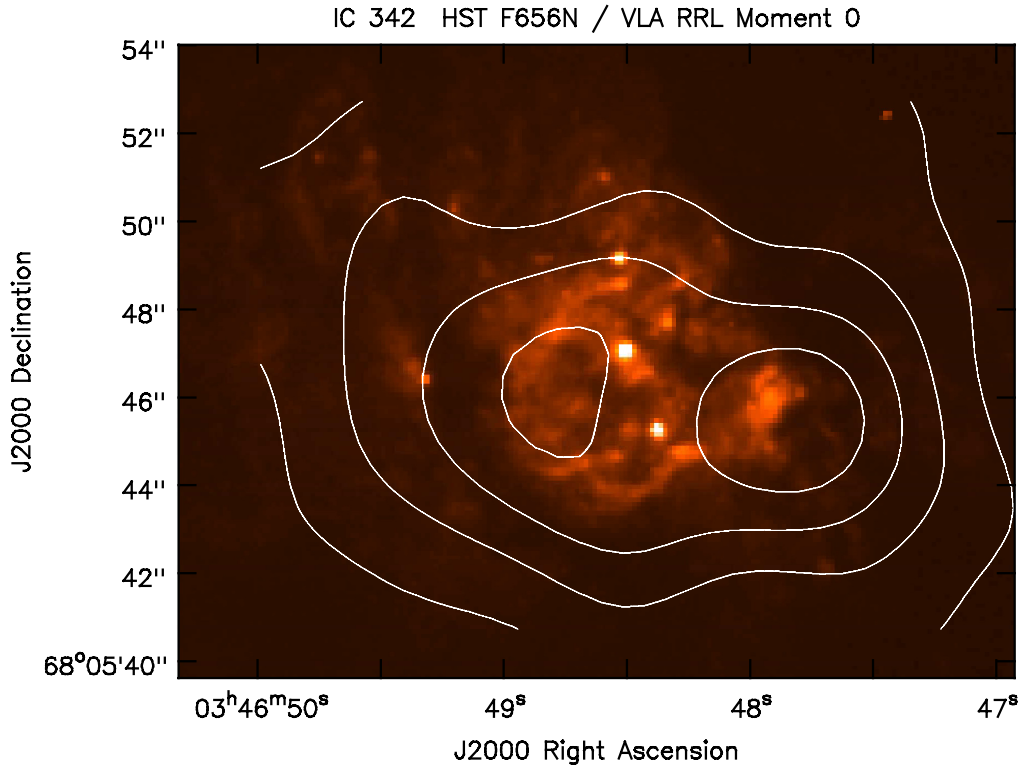} 
\caption{Star formation diagnostics in the \ic{342} nuclear region.
  Shown in color is the HST H$\alpha$ (F656N) image with the JVLA
  Ka-band RRL integrated intensity map in contours.  The contour
  levels are 0.2, 0.4, 0.6, and 0.8 times the peak intensity
  ($0.1324\,$Jy$\,$beam$^{-1}$\kms).  The synthesized beam size, not
  shown for clarity, is $4\arcsper5 \times 4\arcsper5$.}
\label{fig:hst_vla}
\end{figure}

\ic{342} is near the Galactic equator and thus there is significant
obscuration from dust at optical wavelengths, but nevertheless this
galaxy is well studied across the electromagnetic spectrum.  An HST
H$\alpha$ (F656N) image\footnote{Taken from the HST data archive from
  proposal ID 6367 observed on 1996 January 7.  The rms positional
  accuracy of the HST WFPC2 data is $\sim 1\arcsper6$
  \citep{ptak06}. Here we have not attempted to improve the
  astrometric registration since the JVLA resolution is $4\arcsper5$.}
from the Wide Field Planetary Camera 2 (WFPC2) reveals three star
clusters within the central 10\arcsec\ of the nuclear region (see
Figure~\ref{fig:hst_vla}).  Our JVLA RRL integrated intensity map,
overlaid as contours, shows the location and intensity of the thermal
emission.  The peak RRL emission lies to the east and west of the
nuclear star cluster.

{\it Chandra} High Resolution Camera (HRC-I) observations of \ic{342}
resolve X-ray emission in the nucleus into two sources: C12 and C13.
The brighter source (C12) lies close to the central star cluster and
has an extended component with an X-ray luminosity of $6.7 \pm 0.5
\times 10^{38}\,$erg$\,$s$^{-1}$ between $0.08-10\,$kev, consistent
with the expectations of a nuclear starburst \citep{mak08}.  There is
some evidence that \ic{342} contains a radio-quiet active galactic
nuclei based on long term X-ray variability, but this is not
conclusive \citep{mak08}.

\begin{figure}
\includegraphics[angle=0,scale=0.8]{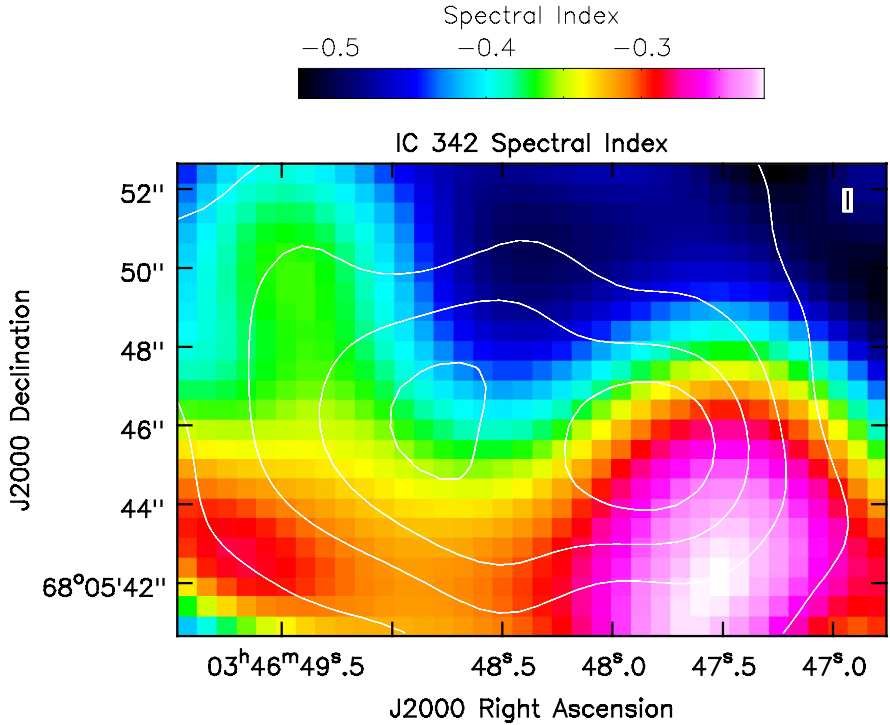} 
\caption{Spectral index map of the nuclear region in \ic{342}.  The
  spectral index ($S_{\rm C} \propto \nu^{\rm s}$) was calculated
  using the 6.7\ghz\ and 35\ghz\ images.  The contours correspond to
  the Ka-band RRL integrated intensity map with levels as in
  Figure~\ref{fig:hst_vla}.  The synthesized beam size, not shown for
  clarity, is $4\arcsper5 \times 4\arcsper5$.}
\label{fig:spix}
\end{figure}

In normal galaxies radio continuum observations trace both thermal
bremsstrahlung emission from \hii\ regions and non-thermal synchrotron
emission from relativistic electrons in supernova remnants and diffuse
gas \citep{condon92}.  These processes are typically distinguished by
observations at several different frequencies to derive the spectral
index ($S_{\rm C} \propto \nu^{\alpha}$), where for thermal emission
$\alpha \sim -0.1$ and for non-thermal emission $\alpha \sim -0.5$ to
$-1.2$ \citep{lisenfeld00}.  Figure~\ref{fig:spix} shows the spectral
index map created using the C-band (6.7\ghz) and Ka-band (35\ghz)
images.  The RRL integrated intensity contour map is overlaid and
indicates the location of thermal emission.  The spectral index varies
between $\alpha \sim -0.2$ (just south of the West RRL component) to
$\alpha \sim -0.5$ (in the northwest).  This generally agrees with
previous work except toward the northeast where we measure $\alpha
\sim -0.35$ compared to $\alpha > 0$ \citep[c.f.][]{turner83,
  meier01}.  We estimate that the errors in our spectral index map are
relatively small over the region shown since the continuum emission is
bright.

The radio continuum morphology at 2.6\mm\ wavelength should primarily
trace \hii\ regions since we expect the non-thermal emission and
thermal dust emission to be significantly weaker at these wavelengths.
Comparison of the 2.6\mm\ and 1.3\mm\ continuum images from
\citet{meier01} with our JVLA RRL integrated intensity map are
generally in good agreement with two main components surrounding the
nuclear star cluster.  The RRL East component, however, is located
south of the millimeter peak.  \citet{meier01} suggest that the
1.3\mm\ emission is dominated by thermal dust emission based on the
spectral index, and that a fraction of the 2.6\mm\ emission may arise
from thermal dust emission.  This may explain the differences between
their millimeter continuum images and our RRL integrated intensity
maps.  It is also possible that the gas toward the northeast is
optically thick at 35\ghz.  JVLA continuum observations of \ic{342} at
cm-wavelengths with higher spatial resolution ($0\arcsper3$) reveal
about a dozen compact sources in the nucleus \citep{tsai06}.  Some of
these components are consistent with compact, dense \hii\ regions as
predicted by our models.

We measure a total continuum flux density of 28.6\mjy\ at 35\ghz\ over
the whole region (see Table~\ref{tab:cont}).  We predict a total flux
density of $\sim 20$\mjy\ at 3\mm\ (see
Figure~\ref{fig:mhm_wholeRegion}), about 25\% less than measured by
\citet{meier01}.  Here the ``whole region'' is defined in terms of the
RRL emission.  The continuum emission extends beyond the RRL emission
(see Figure~\ref{fig:cont}).  The continuum emission defined over the
full extent of the nucleus is 34\mjy\ at 35\ghz, which accounts for
some of this difference.  \citet{rabidoux14} measured a total flux
density of $27.82 \pm 1.1$\mjy\ at 35\ghz\ using the Green Bank
Telescope with a half-power beam-size of 23\arcsec.  Therefore our
JVLA measurements are not missing any significant flux density in the
nuclear region due to the lack of zero spacing data.  Our models
predict a hydrogen-ionizing rate of $N_{\rm L} \sim 2 \times
10^{52}\,$s$^{-1}$ over the whole region, corresponding to $\sim 2000$
O6 stars of luminosity class V \citep{martins05}.  Assuming a solar
metallicity, continuous star formation, and a Kroupa initial mass
function we estimate a SFR $\sim 0.15$\sfr.  This is consistent with
other estimates based on measurements of the thermal radio continuum
emission \citep{meier01, rabidoux14}.

The molecular gas in the nucleus of \ic{342} has been extensively
studied with several tracers: CO and various isotopomers \citep{meier01,
  israel03}; dense gas tracers HCN and HNC \citep{downes92, meier05};
the kinetic temperature probe NH$_{3}$ \citep{montero-castano06,
  lebron11}; and shock tracers SiO and CH$_{3}$OH \citep{meier05,
  usero06}. The dense molecular gas is concentrated into five giant
molecular clouds (GMCs) and forms a mini spiral \citep{downes92}.
\citet{meier05} observed the nucleus with many molecular tracers at
high spatial resolution and proposed the following picture.  The mini
spiral was created by the response of material from the potential of
the bar.  As gas collides at the location of the spiral arms energy is
lost.  This plus tidal torquing allows material to drift inward and
trigger star formation where gas piles up at the intersection of the
arms and central ring.  They suggested trailing spiral arms, where the
tips of the arms point toward the direction of gas motion, since the
molecular gas densities are higher on the leading (counterclockwise)
edges of the spiral arms.  Using higher spatial resolution CO and HCN
data, \citet{schinnerer08} proposed evidence for feedback where star
formation alters the flow of molecular material before reaching the
nucleus, thereby inhibiting star formation.  \citet{rollig16} proposed
a slightly modified picture where the mini spiral is flipped by
90\degree, creating leading spiral arms.  This is based on SOFIA
observations of [\cii]$\,158$\micron\ and [\nii]$\,205$\micron.  The
[\cii]$\,158$\micron\ emission arises from the \hii\ regions and
photodissociation regions (PDRs), whereas the [\nii]$\,205$\micron\
emission only traces \hii\ regions.  They detected redshifted emission
toward the southeast and blueshifted emission toward the northwest,
consistent with leading spiral arms.

\begin{figure}
\includegraphics[angle=0,scale=0.8]{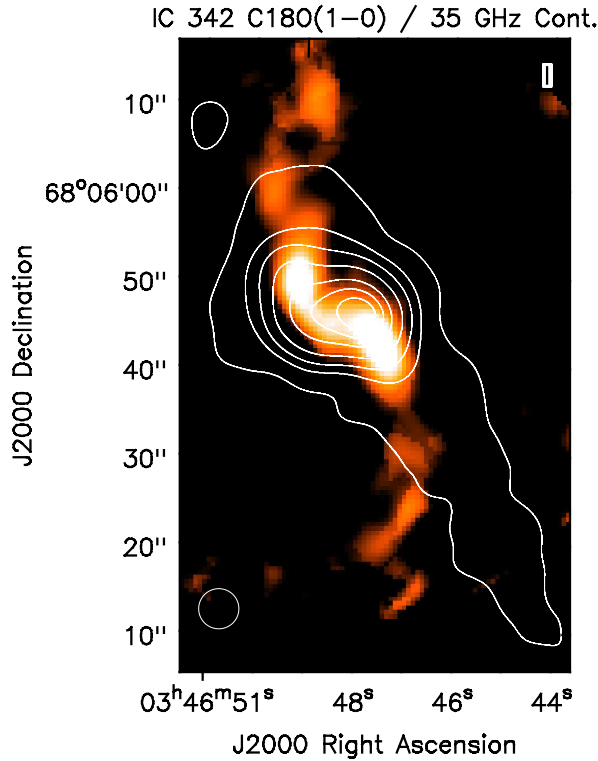} 
\caption{Integrated intensity image of the C$^{18}$O(1-0) transition
  of the central regions of \ic{342} from the Owens Valley Millimeter
  Array \citep{meier01}. The contours are the JVLA Ka-band continuum
  emission shown in Figure~\ref{fig:cont}.  The circle in the bottom
  left-hand corner is the beam size of the JVLA data.}
\label{fig:co}
\end{figure}

The JVLA RRL data probe thermal emission only arising from \hii\
regions.  The RRL East and West components are spatially associated
with GMC C and GMC B, respectively, as defined by \citet{downes92}.
The RRL East component is toward the south of GMC C, or near GMC C3 as
defined by \citet{meier01}.  The molecular and ionized thermal gas
also have similar velocity structure.  The RRL East/GMC C region has
velocities between $V_{\rm bary} = 40-50$\kms, whereas the RRL
west/GMC B region has velocities between $V_{\rm bary} = 10-20$\kms.
We cannot confirm the velocity structure of the ionized gas observed
by \citet{rollig16} since the RRL signal falls off quickly from the
nuclear region (see Figure~\ref{fig:moment}).  We do detect radio
continuum emission, however, along the southern spiral arm which is
located to the west of the molecular material (see
Figure~\ref{fig:co}).  The radio continuum emission is on the leading
edge of the arm relative to the molecular material, consistent with
the trailing arm picture proposed by \citet{meier05}.

\section{Summary}

\ic{342} is a face-on, late-type spiral galaxy with a nuclear region
similar to the center of the Milky Way.  Here we measure the RRL and
continuum emission with the JVLA at C-band (5 and 6.7\ghz) and Ka-band
(33 and 35\ghz), to probe star formation in the nucleus of \ic{342}.
The RRLs consist of \hrrl{106}-\hrrl{113} at 5\ghz;
\hrrl{97}-\hrrl{101} at 6.7\ghz; \hrrl{58} at 34\ghz; and \hrrl{57}
at 35\ghz.  The RRL emission probes thermal emission arising from
\hii\ regions, whereas the continuum emission traces both thermal and
non-thermal processes.  The RRL data allow us to cleanly separate the
thermal and non-thermal emission and also provides kinematic
information.

We detect RRL and continuum emission at all frequencies.  The RRL
emission is concentrated into two components east and west of the
nuclear star cluster.  These \hii\ regions concentrations are
spatially and kinematically associated with the GMCs measured by dense
molecular gas tracers (e.g., HCN).  To increase the RRL
signal-to-noise ratio we average data over three zones: the ``whole
region'' which encompasses all of the RRL emission; the ``East
component'' which isolates the east RRL component; and the ``West
component'' which isolates the west RRL component.

We model these regions in two ways: a simple model consisting of
uniform gas radiating in spontaneous emission (SEM), or as a
collection of many compact \hii\ regions in non-LTE (MHM).  The MHM is
a better fit to the data and predicts many (hundreds), compact ($\lsim
0.1$\pc), dense ($n_{\rm e} \sim 10^{4}-10^{5}$\percc) \hii\ regions.
The models cannot constrain the electron temperature or the number of
\hii\ regions very well, but all models predict similar results for
the other physical properties.  For example, all models that fit the
data require compact, dense \hii\ regions that are consistent with high
spatial resolution radio continuum data \citep[e.g., see][]{tsai06}.
All models predict that the non-thermal emission dominates at 5\ghz,
whereas the thermal emission dominates at 35\ghz.  We estimate a
non-thermal spectral index $\alpha \sim -0.8$.  For the entire nuclear
region we calculate a hydrogen ionizing rate of $N_{\rm L} \sim$
\nexpo{2}{52}$\,{\rm s}^{-1}$, consistent with previous estimates
based on radio continuum observation alone.  This corresponds to
a star formation rate of $\sim 0.15$\sfr.

The RRL data provide kinematic information that in principle can
constrain the dynamics of the nuclear region.  We cannot confirm the
proposal by \citet{rollig16} that the mini molecular spiral is a
leading spiral based on the red/blue-shifted lines of
[\nii]$\,205$\micron\ toward the southeast/northwest.  The RRLs
intensity falls off quickly from the center of the two components.
But we do detect radio continuum emission west of the southern
molecular mini spiral arm, consistent with trailing spiral arms
originally proposed by \citet{meier05}.

\acknowledgments

We thank Dave Meier for providing the C$^{18}$O(1-0) image of
\ic{342}.  This research was supported by NSF grant 1413231
(P.I. K. Johnson).  T.V.W. is supported by the NSF through the Grote
Reber Fellowship Program administered by Associated Universities,
Inc./National Radio Astronomy Observatory, the D.N. Batten Foundation
Fellowship from the Jefferson Scholars Foundation, the Mars Foundation
Fellowship from the Achievement Rewards for College Scientists
Foundation, and the Virginia Space Grant Consortium.

\vspace{5mm}
\facilities{JVLA}

\software{CASA (v4.0.1)}

\appendix

\section{\hii\ Region Models}

Following \citet{puxley91}, we model the central region of \ic{342} in
two diverse ways.  First we consider a very simple model that assumes
the ionized gas has constant density and temperature, and that the
RRLs are formed under LTE with no stimulation emission.  We call this
the spontaneous emission model (SEM).  Second, we assume the ionized
gas is contained in a collection of compact \hii\ regions all with the
same size, density, and temperature.  We call this the multiple \hii\
region model (MHM).  The numerical code was developed in Python and is
available on GitHub \citep{wenger17}.  Below we describe each model in
detail for completeness.

\subsection{Spontaneous Emission Model (SEM)}\label{sec:sem}

Here we follow \citet{puxley91} (also see \citet{bell78}) and express
the total line flux density from a single \hii\ region as
\begin{equation} \label{eq:s_tot}
S_{\rm L} = \Omega_{\rm HII} B(\nu_{\rm L}, T_{\rm e}) 
\left[ \left(\frac{b_{\rm n} \tau_{\rm L}^{*} + \tau_{\rm c}}{\tau_{\rm L} + \tau_{\rm c}} \right)
(1 - e^{-(\tau_{\rm L} + \tau_{\rm c})}) - (1 - e^{-\tau_{\rm c}}) \right] + 
S_{\rm bg} e^{-\tau_{\rm c}} (e^{-\tau_{\rm L}} - 1),
\end{equation}
where $\Omega_{\rm HII}$ is the \hii\ region solid angle, $B(\nu_{\rm
  L}, T_{\rm e})$ is the Planck function, $\nu_{\rm L}$ is the line
frequency, \te\ is the electron temperature, $S_{\rm bg}$ is the
background continuum flux density, and $\tau_{\rm c}$ and $\tau_{\rm
  L}$ are the continuum and line opacity, respectively.  The line
opacity is $\tau_{\rm L} = b_{\rm n} \beta_{\rm n} \tau_{\rm L}^{*}$,
where $b_{\rm n}$ and $\beta_{\rm n}$ are the departure coefficients
of the energy level n, and $\tau_{\rm L}^{*}$ is the line opacity in
LTE.  The contribution from spontaneous emission is then
\begin{equation} \label{eq:s_spon}
S_{\rm L} = \Omega B(\nu_{\rm L}, T_{\rm e}) e^{-\tau_{\rm c}} (1 - e^{-\tau_{\rm L}})
\end{equation}
where in LTE $\tau_{\rm L} = \tau_{\rm L}^{*}$.  The LTE line opacity
for Hn$\alpha$ RRLs is approximated by
\begin{equation} \label{eq:tau_L}
\tau_{\rm L}^{*} = \nexpo{1.7}{3} \left( \frac{\Delta{\nu}_{\rm L}}{\rm kHz} \right)^{-1}
\left( \frac{T_{\rm e}}{\rm K} \right)^{-2.5} 
\left( \frac{EM}{\rm cm^{-6}\, pc} \right), 
\end{equation}
where $EM = \int n_{\rm e}^{2} d\ell$ is the emission measure,
$\Delta{\nu}$ is the line width, \Ne\ is the electron density, and $d\ell$ is 
the path length. The thermal continuum flux density is 
\begin{equation} \label{eq:s_cont}
S_{\rm th} = \Omega B(\nu_{\rm L}, T_{\rm e}) (1 - e^{-\tau_{\rm c}}),
\end{equation}
where the continuum opacity is approximated by
\begin{equation} \label{eq:tau_c}
\tau_{\rm c} = \nexpo{8.2}{-2} \left( \frac{T_{\rm e}}{\rm K} \right)^{-1.35} 
\left( \frac{\nu_{\rm L}}{\rm GHz} \right)^{-2.1}
\left( \frac{EM}{\rm cm^{-6}\, pc} \right).
\end{equation}

\subsection{Multiple H$\,${{\scriptsize \sl II}} Regions Model (MHM)}\label{sec:mhm}

Here we follow \citet{anantharamaiah93} where the RRL emission is
produced by a collection of $N_{\rm HII}$ identical \hii\ regions with
electron density \Ne, electron temperature \te, and linear size
$\ell$, embedded in a larger region of continuum emission.  From
Equation~\ref{eq:s_tot} the total flux density emitted from a single
\hii\ region is
\begin{equation} \label{eq:s_tot2}
S_{\rm L} = \frac{2kT_{\rm e}\nu^{2}}{c^{2}}\Omega_{\rm HII} 
\left[ \left(\frac{b_{\rm n} \tau_{\rm L}^{*} + \tau_{\rm c}}{\tau_{\rm L} + \tau_{\rm c}} \right)
(1 - e^{-(\tau_{\rm L} + \tau_{\rm c})}) - (1 - e^{-\tau_{\rm c}}) \right] + 
\frac{1}{2}\left( \frac{\Omega_{\rm HII}}{\Omega_{\rm L}}\right)
S_{\rm bg} e^{-\tau_{\rm c}} (e^{-\tau_{\rm L}} - 1),
\end{equation}
where $\Omega_{\rm L}$ is the solid angle of the entire emitting
region.  Here we assume the Rayleigh-Jeans limit ($h\nu \ll kT_{\rm
  e}$) and $B(\nu_{\rm L}, T_{\rm e}) = 2kT_{\rm e}\nu^{2}/c^{2}$.
The non-LTE population levels are given by $b_{\rm
  n}$, whereas $\beta_{\rm n}$ is the amplification term given by
\begin{equation} \label{eq:beta}
\beta_{\rm n} = 1 - \frac{k T_{\rm e}}{h \nu}\frac{d(\ln\,b_{\rm n})}{d{\rm n}} \Delta{n}.
\end{equation}
We calculate the departure coefficients using the code developed by
\citet{salem79}, extended by \citet{walmsley90}, and available in
Appendix E.1 of \citet{gordon09}.

For these non-LTE models we use the more accurate expressions for line
and continuum optical depth from \citet{viner79}.  The LTE line
optical depth is
\begin{equation} \label{eq:tau_L2}
\tau_{\rm L}^{*} = 1.4854 \times 10^{-22}\, n_{\rm e} n_{\rm i} n^{2} f_{\rm n,m}
T_{\rm e}^{-2.5}\, exp\left( \frac{1.57803 \times 10^{5}}{n^{2}T_{\rm e}}\right)
\frac{H\,\ell}{\Delta{V_{\rm D}}},
\end{equation}
where $n_{\rm e}$ is the electron density in cm$^{-3}$, $n_{\rm i}$ is
the ion density in cm$^{-3}$, \te\ is the electron temperature in K,
$\ell$ is the path length in cm, and $\Delta{V_{\rm D}}$ is the Doppler
line width in \kms.  For a gas composed of hydrogen and helium
\begin{equation} \label{eq:ni}
n_{\rm i} = n_{\rm e}\left[ 1 + \frac{n_{\rm He^{+}}}{n_{\rm H^{+}}} + 
\frac{2n_{\rm He^{++}}}{n_{\rm H^{+}}}\right] ^{-1}.
\end{equation}
A typical Milky Way value for the singly ionized helium abundance
ratio is $n_{\rm He^{+}}/n_{\rm H^{+}} = 0.07$ \citep{wenger13}.  The
radiation field that ionized Milky Way \hii\ regions is not hard
enough to produce doubly ionized helium and we therefore set $n_{\rm
  He^{++}}/n_{\rm H^{+}} = 0$.  Only in very metal poor \hii\ regions
has doubly ionized helium been detected in significant amounts.  Since
the metallicity of the central regions of \ic{342} is similar to the
Solar metallicity we expect little doubly ionized helium to exist.
The classical oscillator strength is given by \citet{menzel69}
\begin{equation} \label{eq:f}
f_{\rm n,m} = nM_{1}(\Delta{n})
\left[ 1 + \frac{3}{2}\frac{\Delta{n}}{n} + \frac{M_{2}(\Delta{n})}{n^{2}}\right],
\end{equation}
where for $\Delta{n} = 1$, $M_{1} = 0.1907749$ and $M_{2} =
-0.1877268$.  We assume the line is broadened by thermal and
non-thermal Doppler motions and pressure broadening from electron
impacts.  The Doppler line width is given by
\begin{eqnarray} \label{eq:doppler}
\Delta{V_{\rm D}} & = & 2(\ln(2))^{1/2}\left( \frac{2kT_{\rm e}}{m_{\rm H}} +
\frac{2}{3}\langle V_{\rm T} \rangle^{2}\right)^{1/2} \nonumber \\
& = & 1.66511\left( 0.016499\,T_{\rm e} + \frac{2}{3}\langle V_{\rm T} \rangle^{2}\right)^{1/2}
{\rm km \, s}^{-1},
\end{eqnarray}
where $m_{\rm H}$ is the hydrogen mass and the non-thermal Doppler
component is expressed in terms of a turbulent velocity, $V_{\rm T}$.
Here the units of $T_{\rm e}$ are in K and $V_{\rm T}$ are in \kms.
Typical Milky Way \hii\ region values of the turbulent velocity are
$\sim 12$\kms\ \citep{balser11}.  Pressure broadening produces a
Lorentzian line shape with wider wings than the Gaussian profile
produced by pure Doppler motions.  The convolution of a Gaussian and a
Lorentzian produces a Voit profile given by
\begin{equation} \label{eq:voit}
H(a,u) = \frac{a}{\pi}\,\int_{-\infty}^{\infty}
\frac{exp(-y^{2})dy}{a^{2} + (u - y)^{2}},
\end{equation}
where $a = (\ln\,2)^{1/2}\,\Delta{\nu_{\rm L}}/\Delta{\nu_{\rm D}}$
and $u = 2(\ln\,2)^{1/2}(\nu - \nu_{\rm L})/\Delta{\nu_{\rm D}}$.
From the Doppler equation $\Delta{\nu_{\rm D}} = \nu\,\Delta{V_{\rm
    D}}/c$.  For hydrogen the Lorentzian-to-Doppler line width ratio
is approximated as
\begin{equation} \label{eq:ratio}
\frac{\Delta{\nu_{\rm L}}}{\Delta{\nu_{\rm D}}} \approx 9.76 \times 10^{-16}\,
n_{\rm e} T_{\rm e}^{-1/2} T_{\rm D}^{-1/2} \frac{n^{7}}{\Delta{n}}
\left[ -11.4 + \ln \left( T_{\rm e}\frac{n}{\Delta{n}} \right) \right],
\end{equation}
where the Doppler temperature is $T_{\rm D} = T_{\rm e} +
40.42\,\langle V_{\rm T}^{2} \rangle$.  Here the electron density has
units of cm$^{-3}$ and the temperatures are in Kelvin.

The thermal continuum opacity is
\begin{equation} \label{eq:tau_c2}
\tau_{\rm c} = 9.7699 \times 10^{-21}\,
T_{\rm e}^{-1.5}\nu^{-2}n_{\rm e}n_{\rm iT}\,A\,\ell,
\end{equation}
where \te\ is the electron temperature in K, $\nu$ is the frequency in GHz, \Ne\
is the electron density in cm$^{-3}$ , and $\ell$ is the path length in cm.  Also,
\begin{equation} \label{eq:nit}
n_{\rm iT} = n_{\rm e}\left[ 1 + \frac{n_{\rm He^{+}}}{n_{\rm H^{+}}} + 
\frac{4n_{\rm He^{++}}}{n_{\rm H^{+}}}\left(1 - \frac{\ln\,2}{A}\right)\right]
\left[1 + \frac{n_{\rm He^{+}}}{n_{\rm H^{+}}} + 
\frac{2n_{\rm He^{++}}}{n_{\rm H^{+}}}\right]^{-1},
\end{equation}
and $A = \ln(0.04955\,T_{\rm e}^{1.5} / \nu)$. 

The number of \hii\ regions, $N_{\rm HII}$, is calculated by taking
the total observed line flux and dividing by the integrated line flux
for one region ($S_{\rm int}^{\rm HII}$).  Assuming a Gaussian line
profile $S_{\rm int}^{\rm HII} = 1.064\,S_{\rm L}(\nu = \nu_{\rm
  L})\,\Delta{V_{\rm D}}$, where the line width is expressed as the
FWHM.  The thermal continuum flux density is
\begin{eqnarray} \label{eq:thermal}
S_{\rm th} & = & \frac{2k\nu^{2}}{c^{2}}\Omega_{\rm HII} 
T_{\rm e}(1 - e^{-\tau_{\rm c}})N_{\rm HII} \,\,\,\,\,\,\, \tau_{\rm c} \ll 1 \nonumber \\
           & = & \frac{2k\nu^{2}}{c^{2}}\Omega_{\rm L} 
T_{\rm e}(1 - e^{-\tau_{\rm c}^{\rm los}}) \,\,\,\,\,\,\,\,\,\,\,\,\,\,\,\,\,\,\,\,\, \tau_{\rm c} \gsim 1
\end{eqnarray}
where $\tau_{\rm c}^{\rm los} = N_{\rm HII}^{\rm los} \tau_{\rm c}$ and
$N_{\rm HII}^{\rm los} = N_{\rm HII}\frac{\ell^{2}}{L^{2}}$.  Here
$\ell$ is the linear size of an single \hii\ region, whereas $L$ is
the linear size of the line-emitting region.  So if the continuum
optical depth is small we just multiple the flux density of one \hii\
region times the number of \hii\ regions.  As the optical depth
becomes large we approximate the total flux using the optical depth
along the line-of-sight, $\tau_{\rm c}^{\rm los}$.  The non-thermal
flux density is then
\begin{equation} \label{eq:non-thermal}
S_{\rm nth} = S_{\rm tot} - S_{\rm th}
\end{equation}
where $S_{\rm tot}$ is the observed continuum flux density.  But
$S_{\rm nth}$ is the result of an intrinsic non-thermal emission,
$S_{\rm nth0}$, absorbed by free-free processes from the intervening
\hii\ regions or
\begin{equation} \label{eq:non-thermal2}
S_{\rm nth} \approx S_{\rm nth0} \left[ 1 - \frac{1}{2}N_{\rm HII}^{\rm los}\,(1 - e^{-\tau_{\rm c}})\right].
\end{equation}
This assumes a random distribution of \hii\ regions within the
line-emitting region, $L$, and thus that on average every \hii\ region
is located half-way inside this region.  This approximation is valid
for all values of $N_{\rm HII}^{\rm los}$ for $\tau_{\rm c} \ll 1$ and
all values of $\tau_{\rm c}$ if $N_{\rm HII}^{\rm los} < 1$.  Assuming
$S_{\rm nth0} \propto \nu^{\alpha}$ we can estimate a non-thermal
spectral index.

\section{Star Formation Rate}\label{sec:sfr}

We estimate the star formation rate (SFR) using {\it Starburst99}, a
software program designed to model various properties of star forming
galaxies \citep{leitherer99}.  Following \citet{murphy11}, we
calculate the SFR by assuming solar metallicity, continuous star
formation, and a Kroupa initial mass function for stellar masses
between $0.1-100$\msun\ as
\begin{equation} \label{eq:sfr}
SFR = 7.29 \times 10^{-54}  \left( \frac{N_{\rm L}}{\rm s^{-1}}\right) \,\,\,\,\, {\rm M}_{\odot}\,{\rm yr}^{-1},
\end{equation}
where $N_{\rm L}$ is the number of hydrogen-ionizing photons emitted
per second \citep[c.f.,][]{calzetti07}.  For an ionization bounded
nebula the total number of ionizations equals the number of
recombinations.  Assuming a spherical geometry
\begin{equation} \label{eq:nl}
N_{\rm L} = \frac{4\,\pi}{3}R^{3}n_{\rm e}^{2}\alpha_{\rm B},
\end{equation}
where $R$ is the radius and $\alpha_{\rm B}$ is the case B
recombination rate.  \citet{hui97} approximate the recombination
rate by
\begin{equation} \label{eq:alphab}
\alpha_{\rm B} = 2.753 \times 10^{-14}\frac{\lambda_{\rm HI}^{1.5}}
{[1 + (\lambda_{\rm HI}/2.740)^{0.407}]^{2.242}} \,\,\,\,\, {\rm cm}^{-3}\,{\rm s}^{-1},
\end{equation}
where $\lambda_{\rm HI} = 315614/T_{\rm e}$.  This approximation is a
fit to the data in \citet{ferland92} and is accurate to between 0.7\%
and 1\% for temperatures up to $10^{9}\,$K.

For the SEM we use the line-emitting region size to calculate the
radius (i.e., $R = L/2$), whereas for the MHM we use the single \hii\
regions size (i.e., $R = \ell/2$).  For the MHM we multiply $N_{\rm L}$
for a single \hii\ region by the number of \hii\ regions, $N_{\rm
  HII}$, to calculate the total number of hydrogen-ionizing photons
emitted per second within the region modeled.

{}


\begin{thebibliography}{}

\bibitem[Ambartsumian (1954)]{ambart54} Ambartsumian, V.A. 1954,
  I.A.U. Transactions, 8, 665

\bibitem[Anantharamaiah \& Goss (1996)]{anantharamaiah96}
  Anantharamaiah, K. R., \& Goss, W. M. 1996, \apj, 466, L13

\bibitem[Anantharamaiah et al.(2000)]{anantharamaiah00}
  Anantharamaiah, K. R., Viallefond, F., Mohan, N. R., Goss, W. M., \&
  Zhao, J.-H. 2000, \apj, 537, 613

\bibitem[Anantharamaiah et al.(1993)]{anantharamaiah93}
  Anantharamaiah, K. R., Zhao, J.-H., Goss, W. M., \& Viallefond,
  F. 1993, \apj, 419, 585

\bibitem[Babu \& Feigelson(2006)]{babu06} Babu, G. J., \& Feigelson,
  E. D. 2006, in ASP Conf. Ser. 351, Astronomical Data Analysis
  Software and Systems XV, ed. C. Gabriel, C. Arviset, D. Ponz, \&
  S. Enrique (San Francisco, CA: ASP), 127

\bibitem[Balser et al.(1999)]{balser99} Balser, D. S., Bania, T. M.,
  Rood, R. T., \& Wilson, T. L. 1999, \apj, 510, 759

\bibitem[Balser et al.(2011)]{balser11} Balser, D. S., Rood, R. T,
  Bania, T. M., \& Anderson, L. D. 2011, \apj, 738, 27

\bibitem[Balser et al.(2016)]{balser16} Balser, D. S., Roshi, D. A.,
  Jeyakumar, S., et al. 2016, \apj, 824, 125

\bibitem[Bell \& Seaquist(1978)]{bell78} Bell, M. B., \& Seaquist,
  E. R. 1978, \apj, 223, 378

\bibitem[B\"{o}ker et al. (2004)]{boker04} B\"{o}ker, T., Walcher,
  C.-J., Rix, H.-W., et al. 2004, in ASP Conf. Ser. 322, The Formation
  and Evolution of Massive Young Star Clusters, ed. H.J.G.L.M. Lamers,
  L.J. Smith, and A. Nota (San Francisco: ASP), 39

\bibitem[Calzetti et al.(2007)]{calzetti07} Calzetti, D., Kennicutt,
R. C., Engelbracht, C. W. et al. 2007, \apj, 666, 870

\bibitem[Condon(1992)]{condon92} Condon, J. J. 1992, \araa, 30, 575

\bibitem[de Grijs (2004)]{degrijs04} de Grijs, R. 2004, in ASP
  Conf. Ser. 322, The Formation and Evolution of Massive Young Star
  Clusters, ed. H.J.G.L.M. Lamers, L.J. Smith, and A. Nota (San
  Francisco: ASP), 29

\bibitem[Downes et al.(1992)]{downes92} Downes, D., Radford, S. J. E.,
  Guilloteau, S., et al. 1992, \aap, 262, 424

\bibitem[Ferland et al.(1992)]{ferland92} Ferland, G. J., Peterson,
  B. M., Horne, K., Welsh, W. F., \& Nahar, S. N.  1992, \apj, 387, 95

\bibitem[Gordon \& Sorochenko(2009)]{gordon09} Gordon, M. A., \&
  Sorochenko, R. L. 2009, Astronomy and Space Science Library,
  Vol. 282, Radio Recombination Lines

\bibitem[Greisen(2012)]{greisen12} Greisen, E. 2012, AIPS Cookbook

\bibitem[Holtzman, et al. (1992)]{holtzman92} Holtzman, C. et
  al. 1988, \aap, 201, 23

\bibitem[Hui \& Gnedin(1997)]{hui97} Hui, L., \& Gnedin, N. Y. 1997,
  \mnras, 292, 27

\bibitem[Israel \& Baas(2003)]{israel03} Israel, F. P., \& Baas,
  F. 2003, \aap, 404, 495

\bibitem[Johnson \& Conti(2000)]{johnson00} Johnson, K. E., \& Conti,
  P. S. 2000, \aj, 119, 2146

\bibitem[Johnson, et al.(2004)]{johnson04} Johnson, K. E., Indebetouw,
  R., Watson, C., \& Kobulnicky, H. A. 2004, \aj, 128, 610

\bibitem[Johnson \& Kobulnicky (2003)]{johnson03} Johnson, K. E. \&
  Kobulnicky, H. A. 2003, \apj, 597, 923

\bibitem[Kepley et al.(2011)]{kepley11} Kepley, A. A., Chomiuk, L.,
  Johnson, K. E., et al. 2011, \apjl, 739, L24

\bibitem[Kobulnicky \& Johnson (1999)]{kobulnicky99} Kobulnicky,
  H. A., \& Johnson, K. E. 1999, \apj, 527, 154

\bibitem[Knierman et al. (2004)]{knierman04} Knierman, K., Gallagher,
  S., Charlton, et al. 2004, in ASP Conf. Ser. 322, The Formation and
  Evolution of Massive Young Star Clusters, ed. H.J.G.L.M. Lamers,
  L.J. Smith, and A. Nota (San Francisco: ASP), 59

\bibitem[Larsen (2004)]{larsen04} Larsen, S.S. 2004, in ASP
  Conf. Ser. 322, The Formation and Evolution of Massive Young Star
  Clusters, ed. H.J.G.L.M. Lamers, L.J. Smith, and A. Nota (San
  Francisco: ASP), 19

\bibitem[Lebr\'{o}n et al.(2011)]{lebron11} Lebr\'{o}n, M., Mangum,
  J. G., Mauersberger, R., et al. 2011, \aap, 534, 56

\bibitem[Leitherer (2003)]{leitherer03} Leitherer, C. 2003, in A
  Decade of Hubble Space Telescope Science, ed. M. Livo, K. Noll, \&
  M. Stiavelli (Cambridge: Cambridge Univ. Press), 179

\bibitem[Leitherer et al.(1999)]{leitherer99} Leitherer, C., Schaerer,
  D., Goldader, J. D. et al. 1999, \apjs, 123, 3

\bibitem[Lisenfeld \& V{\"o}lk(2000)]{lisenfeld00} Lisenfeld, U., \&
  V{\"o}lk, H. J. 2000, \aap, 354, 423

\bibitem[Mak et al.(2008)]{mak08} Mak, D. S. Y., Pun, C. S. J., \&
  Kong, A. K. H. 2008, \apj, 686, 995

\bibitem[Martins et al.(2005)]{martins05} Martins, F., Schaerer, D.,
  \& Hillier, D. J. 2005, \aap, 436, 1049

\bibitem[Meier(2014)]{meier14} Meier, D. S. 2014, in IAU Symp. 303,
  eds. L. O. Sjouwerman, C. C. Lang, \& J. Ott, 66

\bibitem[Meier \& Turner(2001)]{meier01} Meier, D. S., \& Turner,
  J. L. 2001, \apj, 551, 687

\bibitem[Meier \& Turner(2005)]{meier05} Meier, D. S., \& Turner,
  J. L. 2005, \apj, 618, 259

\bibitem[Menzel(1969)]{menzel69} Menzel, D. H. 1969, \apjs, 18, 221


\bibitem[Mohan et al.(2002)]{mohan02} Mohan, N. R., Anantharamaiah,
  K. R., \& Goss, W. M. 2002, \apj, 574, 701

\bibitem[Montero-Casta\~{n}o et al.(2006)]{montero-castano06}
    Montero-Casta\~{n}o, M., Hernstein, R. M., \& Ho, P. T. P. 2006, \apj,
    646, 919

\bibitem[Murphy et al.(2011)]{murphy11} Murphy, E. J., Condon, J. J.,
  Schinnerer, E. et al. 2011, \apj, 737, 67

\bibitem[O'Connell (2004)]{oconnell04} O'Connell, R.W. 2004, in ASP
  Conf. Ser. 322, The Formation and Evolution of Massive Young Star
  Clusters, ed. H.J.G.L.M. Lamers, L.J. Smith, and A. Nota (San
  Francisco: ASP), 551

\bibitem[Perley \& Butler(2013)]{perley13} Perley, R. A., \& Butler,
  B. J. 2013, \apjs, 2014, 19

\bibitem[Pilyugin et al.(2014)]{pilyugin14} Pilyugin, L. S., Grebel1,
  E. K., \&Y. Kniazev, A. Y. 2014, \apj, 147, 131

\bibitem[Ptak et al.(2006)]{ptak06} Ptak, A., Colbert, E., van der
  Marel, R. P., et al. 2006, \apjs, 166, 154

\bibitem[Puxley et al.(1991)]{puxley91} Puxley, P. J., Brand,
  P. W. J. L., Moore, T. J. T., Mountain, C. M., \& Nakai, N. 1991,
  \mnras, 585

\bibitem[Puxley et al.(1997)]{puxley97} Puxley, P. J., Mountain,
  C. M., Brand, P. W. J. L., Moore, T. J. T., \& Nakai, N.  1997,
  \apj, 485, 143

\bibitem[Rabidoux et al.(2014)]{rabidoux14} Rabidoux, K., Pisano,
  D. J., Kepley, A. A., Johnson, K. E., \& Balser, D. S.  2014, \apj,
  780, 19

\bibitem[R\"{o}llig et al.(2016)]{rollig16} R\"{o}llig, M., Simon, R.,
  G\"{u}sten, R., et al. 2016, \aap, 591, 33

\bibitem[Rodr\'{i}guez-Rico et al. (2005)]{rodriguez05}
  Rodr\'{i}guez-Rico, C.A., Goss, W.M., Viallefond, F., et al. 2005,
  \apj, 633, 198

\bibitem[Rubin(1985)]{rubin85} Rubin, R. H. 1985, \apjs, 57, 349

\bibitem[Saha et al.(2002)]{saha02} Saha, A., Claver, J., \& Hoessel,
  J. G. 2002, \aj, 124, 839

\bibitem[Salem \& Brocklehurst(1979)]{salem79} Salem, M., \&
  Brocklehurst, M. 1979, \apjs, 39, 633

\bibitem[Schinnerer et al.(2008)]{schinnerer08} Schinnerer, E.,
  B\"{o}ker, T., Meier, D. S., \& Calzetti 2008, \apj, 684, L21

\bibitem[Shaver(1975)]{shaver75} Shaver, P. A. 1975, Prama, 5, 1

\bibitem[Shaver et al.(1983)]{shaver83} Shaver, P. A., McGee, R. X.,
  Newton, L. M., Danks, A. C., \& Pottasch, S. R.  1983, \mnras, 204,
  53

\bibitem[Terlevich (2004)]{terlevich04} Terlevich, R. 2004, in ASP
  Conf. Ser. 322, The Formation and Evolution of Massive Young Star
  Clusters, ed. H.J.G.L.M. Lamers, L.J. Smith, and A. Nota (San
  Francisco: ASP), 11

\bibitem[Tsai et al.(2006)]{tsai06} Tsai, C.-W., Turner, J. L., Beck,
  S. C., et al. 2006, \aj, 132, 2383

\bibitem[Turner et al.(2000)]{turner00} Turner, J. L., Beck, S. C., \&
  Ho, P. T. P. 2000, \aj, 120, 244

\bibitem[Turner \& Ho (1983)]{turner83} Turner, J.L. \& Ho,
  P.T.P. 1983, \apj, 268, L79

\bibitem[Turner \& Ho (1994)]{turner94} Turner, J.L. \& Ho,
  P.T.P. 1994, \apj, 421, 122

\bibitem[Usero et al.(2006)]{usero06} Usero, A., Garc\'{i}a-Burillo,
  S., Mart\'{i}n-Pintado, J., Fuente, A., \& Neri, R. 2006, \aap, 448,
  457

\bibitem[Viner et al.(1979)]{viner79} Viner, M. R., Vall\'{e}e,
  J. P., \& Hughes, V. A. 1979, \apjs, 39, 405

\bibitem[Walmsley(1990)]{walmsley90} Walmsley, C. M. 1990, \aaps, 82,
  201

\bibitem[Wenger (2017)]{wenger17}
Wenger, T. V. 2017, Astrophysics Source Code Library

\bibitem[Wenger et al.(2013)]{wenger13} Wenger, T. V., Bania, T. M.,
  Balser, D. S., \& Anderson, L. D. 2013, \apj, 764, 34

\bibitem[Whitmore (2003)]{whitmore03} Whitmore, B.C. 2003, in A Decade
  of Hubble Space Telescope Science, ed. M. Livo, K. Noll, \&
  M. Stiavelli (Cambridge: Cambridge Univ. Press), 153


\end{thebibliography}
\end{document}